%% file: ex_article.tex
\documentclass[review]{siamart1116}

\input{ex_shared}

\ifpdf
\hypersetup{
  pdftitle={\TheTitle},
  pdfauthor={\TheAuthors}
}
\fi

\usepackage{epstopdf}
\begin{document}

\maketitle
\begin{abstract}
This paper introduces the notion of weak rigidity to characterize a framework by pairwise inner products of inter-agent displacements. Compared to distance-based rigidity, weak rigidity requires fewer constrained edges in the graph to determine a geometric shape in an arbitrarily dimensional space. A necessary and sufficient graphical condition for infinitesimal weak rigidity of planar frameworks is derived. As an application of the proposed weak rigidity theory, a gradient based control law and a non-gradient based control law are designed for a group of single-integrator modeled agents to stabilize a desired formation shape, respectively. Using the gradient  control law, we prove that an infinitesimally weakly rigid formation is locally exponentially stable. In particular, if the number of agents is one greater than the dimension of the space, a minimally infinitesimally weakly rigid formation is almost globally asymptotically stable. In the literature of rigid formation, the sensing graph is always required to be rigid. Using the non-gradient control law based on weak rigidity theory, the sensing graph is unnecessary to be rigid for local exponential stability of the formation.  A numerical simulation is performed for illustrating effectiveness of our main results.
\end{abstract}



\section{Introduction} There is a rapidly growing interest in the study of distributed coordination of networked multi-agent systems due to their wide applications and diverse mathematical challenges. As one of the most significant and challenging problems, the formation stabilization problem, which is concerned with the stabilization of a group of agents via local information to form a desired formation shape, has been studied in a vast amount of references, see, e.g., the survey papers  \cite{Saber07,Ren07,Anderson08,Oh15}.

In the formation stabilization problem, the formation shape is often characterized by specified constraints on agents' states. These constraints differ depending on the sensing graph which describes interaction relationships between agents and sensing capability of agents. In recent years, due to  their advantages in alleviation of computational burden and enhancement of reliability, decentralized formation stabilization strategies based on relative displacement information have received a lot of attention \cite{Fax04,Saber07,Ren07,Xiao09}. Control algorithms proposed in these references often guarantee global stability of the formation but, unfortunately, are  at the cost of being  implemented under a common coordinate system, which is often unavailable when the Global Positioning System (GPS) is disabled. \cite{Oh14b} and \cite{Lee16} proposed two distributed orientation alignment laws for agents to reach agreement on their local coordinate systems, which can efficiently solve the formation problem in the plane in absence of a common  orientation. However, their approaches require all agents have capability of communicating with each other, thus will be invalid when agents equip no communication sensors. Unlike these investigations, the author in \cite{Cortes09} obtained global stability of formation with a discrete-time algorithm via initial orientation alignment. Another hot issue in distributed formation control is bearing-constrained formation. For example, in \cite{Bishop11,Zhao16}, the authors studied how to encode the desired formation by bearing-only constrains. However, the proposed methods also require either the global coordinate system or an orientation synchronization law based on inter-agent communications. \cite{Lin14} obtained a necessary and sufficient condition for global stability of the formation by a consensus-like control protocol, which allows agents to use relative displacements measured in their local coordinate frames, but the stabilizing gain matrix should be designed via a centralized approach.

In order to achieve distributed and communication-free formation in GPS-denied environments, intensive research efforts have been expended. Distance-based formation control, which determines the desired formation by inter-agent distance, and only requires each agent to sense local relative displacements with its local coordinate frame, was investigated in \cite{Anderson08,Krick09,Yu09,Summers09,Summers11,Oh11,Oh14a,Zelazo15,Sun15,Mou16,Sun16auto,Sun16scl,Jing16,Chen17}. By embedding the formation graph into a specified space, a framework consisting of a formation topology and certain coordinates of all vertices is employed to describe the desired formation shape. To solve the formation problem, one should answer the question that how many distance constraints are required to determine the formation framework, which turns out to be equivalent to an Euclidean Distance matrix completion problem \cite{Laurent97,Liberti14,Singer10}. Moreover, this question is also relevant to  network localization problems \cite{Aspnes06}. In the literature, graph rigidity theory \cite{Asimow78,Hendrickson92,Connelly05} was often employed to answer this question and many interesting results have been obtained. In \cite{Krick09}, the authors proposed a gradient based control law for multiple autonomous agents to restore infinitesimally rigid formations under small perturbations from the desired formation shape. In \cite{Yu09} and \cite{Summers11}, the authors solved minimally persistent formation problems under a directed sensing graph by introducing an appropriately designed gain matrix. In \cite{Oh14a}, the authors showed that rigidity of the formation framework is sufficient to ensure local stability of the desired formation. Besides the distance-based formation strategy, \cite{Aranda16} proposed a displacement-based approach to achieving local and global stability of rigid planar formations under different graphs. \cite{Lin16} introduced an affine formation strategy and obtained global stability of formation under universally rigid graphs. \cite{Mou15} proposed a control strategy based on Henneberg vertex additions to achieve a minimally rigid acyclic formation. All these investigations require the target formation shape to be rigid. However, this restriction is not easy to be satisfied in practice due to the demand for a large number of edges in the formation graph.

This paper aims to reduce the number of edges in a graph for determining an undirected formation framework in an arbitrary dimensional space. The fundamental method we propose is based on a modification of the rigidity function in graph rigidity theory. More specifically, we regard pairwise inner products of relative displacements as components of the rigidity function, which are actually constraints determining the desired formation shape. Accordingly, a generalized notion of rigidity, the {\it weak rigidity}, is introduced. Since a distance constraint is equal to the inner product of two identical displacements, weak rigidity can reduce to distance-based rigidity. In fact, one can intuitively observe that angles subtended at vertices are also helpful for determining a desired formation; unfortunately, this information is not efficiently utilized in distance-based formation control. As pointed out in \cite{Oh14b}, the angle information contained in the displacement measurements is difficult to be directly utilized. In this paper, by employing inner products of relative displacements as constraints, angles subtended in the formation graph can be used to determine the desired formation shape. Moreover, the inner product of two vectors in any local coordinate frame is invariant, thus is independent of the global coordinate system. As a result, weak rigidity requires less number of edges than distance-based rigidity to recognize a framework and provides a novel insight to decentralized formation controller synthesis in GPS-denied environments.

The main contributions of this paper are summarized as follows. (i) We define a generalized concept of rigidity called weak rigidity (Definition \ref{def:weak rigidity}), by which a framework in an arbitrarily dimensional space can be determined with fewer constrained edges than distance-based rigidity. In Theorem \ref{rigidtogrigid}, we prove that weak rigidity is necessary but not sufficient for distance-based rigidity, thus is a weaker condition for determining a framework. (ii) For frameworks embedded in the plane, a necessary and sufficient graphical condition is derived for infinitesimal weak rigidity (Theorem \ref{th ns condition for GIS}). It is shown that a framework is infinitesimally weakly rigid if and only if the graph is connected and for each vertex with more than two neighbors, the edges connected to this vertex are not all collinear. Based on the graphical condition, we present two algorithms for constructing a constraint set with minimal number of elements for determining weak rigidity of the framework. (iii) From a matrix completion perspective, we show that by employing weak rigidity theory, the realization problem of a framework is equivalent to a positive semi-definite (PSD) matrix completion problem \cite{Laurent97,Singer10,Liberti14}. Once the PSD matrix is completed, the framework can be uniquely determined up to translations, rotations and reflections. See Theorem \ref{th D=E} and Remark \ref{re congruence}. (iv) We show in Subsection \ref{subsec:generic property} that both weak rigidity and infinitesimal weak rigidity are generic properties of graphs. More precisely, after fixing the graph, either all the frameworks with generic configurations are infinitesimally weakly rigid, or none of them are. (v) As an application, on the basis of weak rigidity theory proposed, we present a gradient based control law for multiple autonomous agents to achieve a desired formation. It is shown that if the number of agents is one greater than the dimension of the space, then almost global asymptotic stability\footnote{A shape is said to be almost globally asymptotically stable if it is asymptotically stable for almost all the initial conditions. That is, the initial conditions converging to incorrect shapes belong to a set of measure zero \cite{Summers09}.} of the minimally infinitesimally weakly rigid formation and collision avoidance can be ensured (Theorem \ref{th n=d+1}). Otherwise the infinitesimally weakly rigid formation is locally exponentially stable (Theorem \ref{th exponential stable}). (vi) A non-gradient based protocol is also proposed for achieving weakly rigid formation. It is shown that once a control gain matrix is properly designed, our control strategy can drive agents to form a locally exponentially stable weakly rigid formation, while the underlying sensing graph is only required to be infinitesimally weakly rigid rather than rigid. This is a relaxed condition for sensing graphs compared to \cite{Krick09,Oh14a,Zelazo15,Sun15,Sun16auto,Sun16scl,Aranda16,Lin16,Chen17}.

This paper is structured as follows. Section \ref{sec:pre} provides preliminaries of graph rigidity theory and center manifold theory. Section \ref{sec:pse} presents the weak rigidity theory. As an application, Section \ref{sec:app}  discusses two control strategies for formation stabilization control and the corresponding stability analysis. Section \ref{sec:simu} presents a  numerical example. Finally, Section \ref{sec:con} concludes the whole paper.

{\it Notations:} Throughout this paper, $\mathbb{R}$ denotes the set of real numbers; $\mathbb{R}^n$ is the $n-$dimensional Euclidean space; $||\cdot||$ stands for the Euclidean norm; $X^T$ means the transpose of matrix $X$; $\otimes$ is the Kronecker product; $\textrm{range}(X)$, $\textrm{null}(X)$ and $\textrm{rank}(X)$ denote the range space, null space, and the rank of matrix $X$; $A\setminus B$ is the set of those elements of $A$ not belonging to $B$; $I_n$ represents the $n\times n$ identity matrix; $\mathbf{1}_n\in\mathbb{R}^{n\times1}$ is a vector with each component being $1$; $\lambda(X)$ is the set of eigenvalues of matrix $X$.

An undirected graph with $n$ vertices and $m$ edges is denoted as $\mathcal{G}=(\mathcal{V},\mathcal{E})$, where $\mathcal{V}=\{1,\cdots,n\}$ and $\mathcal{E}\subseteq\mathcal{V}\times\mathcal{V}$ denote the vertex set and the edge set, respectively. As the graph considered in this paper is undirected, we will not distinguish $(i,j)$ and $(j,i)$.  The incidence matrix is represented by $H=[h_{ij}]$, which is a matrix with rows and columns indexed by edges and vertices of $\mathcal{G}$ with an orientation. $h_{ij}=1$ if the $i$th edge sinks at vertex $j$, $h_{ij}=-1$ if the $i$th edge leaves vertex $j$, and $h_{ij}=0$ otherwise. It is well-known that $\textrm{rank}(H)=n-1$ if and only if graph $\mathcal{G}$ is connected.


\section{Preliminaries}
\label{sec:pre}

\subsection{Graph rigidity theory}
A graph $\mathcal{G}=(\mathcal{V},\mathcal{E})$ can be embedded in $\mathbb{R}^{d}$ by an assignment of locations $p_i\in\mathbb{R}^d$, $i\in\mathcal{V}$ to the vertices. Graph rigidity theory is for answering whether partial length-constrained edges of graph $\mathcal{G}$ can determine the coordinates of the points $p_1,\cdots,p_n$ uniquely up to rigid transformations (translations, rotations, reflections). Several basic definitions related to graph rigidity taken from \cite{Asimow78} and \cite{Hendrickson92} are stated as follows.

The vector $p=(p_1^T,\cdots,p_n^T)^T\in\mathbb{R}^{nd}$ is called a realization or configuration of $\mathcal{G}$. The pair $(\mathcal{G},p)$ is said to be a framework. The {\it rigidity function} $g_{\mathcal{G}}(\cdot):\mathbb{R}^{nd}\rightarrow\mathbb{R}^m$ associated with the framework $(\mathcal{G},p)$ is defined as
\begin{equation}\label{origidfun}
g_{\mathcal{G}}(p)=(\cdots,||e_{ij}||^2,\cdots)^T, ~~(i,j)\in\mathcal{E},
\end{equation}
where $n=|\mathcal{V}|$, $m=|\mathcal{E}|$, $e_{ij}=p_i-p_j$, and $\|e_{ij}\|$ is the Euclidean distance between the vertices $i$ and $j$.

We say two frameworks $(\mathcal{G},p)$ and $(\mathcal{G},q)$ are {\it equivalent} if $g_{\mathcal{G}}(p)=g_{\mathcal{G}}(q)$, i.e., $||p_i-p_j||=||q_i-q_j||$ for all $(i,j)\in\mathcal{E}$. They are {\it congruent} if $||p_i-p_j||=||q_i-q_j||$ for all $i,j\in\mathcal{V}$. A framework $(\mathcal{G},p)$ is called {\it rigid} if there exists a neighborhood $U_p$ of $p$ such that for any $q\in U_p$, once $(\mathcal{G},p)$ is equivalent to $(\mathcal{G},q)$, then they are congruent. $(\mathcal{G},p)$ is {\it globally rigid} in $\mathbb{R}^d$ if it is rigid with $U_p=\mathbb{R}^{nd}$. $(\mathcal{G},p)$ is {\it minimally rigid} if no edges of $\mathcal{G}$ can be removed without losing rigidity of $(\mathcal{G},p)$. For example, the framework in Fig. \ref{fig 4 examples} (a) is both minimally and globally rigid, the framework in Fig. \ref{fig 4 examples} (c) is minimally rigid, while the frameworks in Fig. \ref{fig 4 examples} (b) and Fig. \ref{fig 4 examples} (d) are both non-rigid.

The rigidity function $g_{\mathcal{G}}(p)$ is the key to recognize the framework $(\mathcal{G},p)$. For a time-varying framework, an assignment of velocities that guarantees the invariance of $g_{\mathcal{G}}(p)$, i.e., $\dot{g}_{\mathcal{G}}(p)=0$, is called an {\it infinitesimal motion}. That is,
\begin{equation}\label{infinitesimal}
(v_i-v_j)^T e_{ij}=0~~(i,j)\in\mathcal{E},
\end{equation}
where $v_i=\dot{p}_i$ is the velocity of vertex $i$. Note that rotations, translations, and their combinations always satisfy equation (\ref{infinitesimal}). Such motions are said to be {\it trivial}. A framework is {\it infinitesimally rigid} if every infinitesimal motion is trivial. In a $d$ dimensional space, there are $d$ independent translations and $d(d-1)/2$ independent rotations. Therefore, for a framework $(\mathcal{G},p)$ with $n\geq d$, the dimension of the space formed by trivial motions is $T(n,d)=d+d(d-1)/2=d(d+1)/2$. In fact, equation (\ref{infinitesimal}) is equivalent to $\dot{g}_{\mathcal{G}}(p)=R(p)\dot{p}=0$, where $R(p)\triangleq \frac{\partial g_{\mathcal{G}}(p)}{\partial p}\in\mathbb{R}^{m\times nd}$ is called the {\it rigidity matrix}. Thus one can obtain that a framework $(\mathcal{G},p)$ is infinitesimally rigid if $\textrm{rank}(R(p))=nd-T(n,d)$.

Two vertices of an infinitesimally rigid framework usually do not share identical positions. We present the following lemma to show a necessity condition for existence of overlaps in an infinitesimally rigid framework.
\begin{lemma}\label{le collision avoidance}
Let $(\mathcal{G},p)$ be infinitesimally rigid in $\mathbb{R}^d$. If there exists a vertex $i$ colliding with another vertex, then vertex $i$ has at least $d$ neighbors not colliding with it.
\end{lemma}
\begin{proof} Consider a framework $(\tilde{\mathcal{G}},\tilde{p})$, which is induced by deleting vertex $i$ and all edges involving $i$ from $(\mathcal{G},p)$. Let $\tilde{R}(\tilde{p})$ be the rigidity matrix of $(\tilde{\mathcal{G}},\tilde{p})$. It is easy to see that $\textrm{rank}(\tilde{R}(\tilde{p}))\leq (n-1)d-d(d+1)/2$. Note that when $i$ is added into $(\tilde{\mathcal{G}},\tilde{p})$, the corresponding rigidity function can be written as $g_{\mathcal{G}}=(g_{\tilde{\mathcal{G}}}^T, g^{iT})^T$, where $g_{\tilde{\mathcal{G}}}$ is the rigidity function of $(\tilde{\mathcal{G}},\tilde{p})$, $g^i=(\cdots,||e_{ij}||^2,\cdots)^T$, $j\in\mathcal{N}_i$. Hence, the rigidity matrix of $(\mathcal{G},p)$ is
\[R(p)=\frac{\partial g_\mathcal{G}}{\partial p}=(\frac{\partial g_\mathcal{G}}{\partial \tilde{p}}, \frac{\partial g_\mathcal{G}}{\partial p_i})=\begin{pmatrix}
\tilde{R}(\tilde{p}) & \mathbf{0} \\
\frac{\partial g^i}{\partial \tilde{p}} & \frac{\partial g^i}{\partial p_i}
\end{pmatrix}.\]
Infinitesimal rigidity of $(\mathcal{G},p)$ implies that $\textrm{rank}(\frac{\partial g^i}{\partial p})\geq \textrm{rank}(R(p))-\textrm{rank}(\tilde{R}(\tilde{p}))\geq d$. Therefore $g^i$ should have at least $d$ nonzero components. That is, there exist $N\geq d$ vertices $k_1,\cdots,k_N\in\mathcal{N}_i$ and $p_i\neq p_{k_j}$, $j\in\{1,\cdots,N\}$.
\end{proof}

Finally, it is worth noting that an infinitesimally rigid framework may have or not have overlapped vertices.

\subsection{Center Manifold Theory}

Center manifold theory is a tool of great utility in studying stability of nonhyperbolic equilibria of a nonlinear system. The details of center manifold theory can be found in \cite{Carr12}. Here we introduce a result  for systems with an equilibrium manifold derived in \cite{Summers11}, which will be employed to study stability of equilibria of the formation system.

\begin{lemma}\cite{Summers11}\label{le manifold theory}
Consider the nonlinear autonomous system
\begin{equation}\label{nonlinear system}
\dot{x}=f(x),~~x\in\mathbb{R}^n
\end{equation}
where $f$ is twice continuously differentiable almost everywhere in a neighborhood of the origin. Suppose there exists a smooth $m$-dimensional $(m>0)$ manifold of the equilibrium set $M_1$ for (\ref{nonlinear system}) that contains the origin. If the Jacobian of $f$ at the origin has $m$ eigenvalues with zero real part and $n-m$ eigenvalues with negative real part, then $M_1$ is a center manifold for (\ref{nonlinear system}). Moreover, there exist compact neighborhoods $\Omega_1$ and $\Omega_2$ of the origin such that $M_2=\Omega_2\cap M_1$ is locally exponentially stable, and for each $x(0)\in\Omega_1$, it holds that $\lim_{t\rightarrow\infty}x(t)=q$ for some $q\in M_2$.
\end{lemma}

\section{Weak rigidity}
\label{sec:pse}

Generally speaking, a multi-agent formation problem is ab- out the stabilization of a desired geometric shape formed by a group of mobile agents in a $d-$dimensional space. In the literature, a distance-based formation strategy is often adopted since the global coordinate system is often absent for each agent. In addition to distances, the subtended angles are also available in determining the desired formation shape and are independent of the global coordinate system. The purpose of this paper is to show how to utilize such information in a multi-agent formation problem. In this section, we present a novel approach for recognizing a framework. With the aid of subtended angles information, we show that the total number of edges for recognizing a framework can be reduced.

\begin{figure}
\centering
\includegraphics[width=13cm]{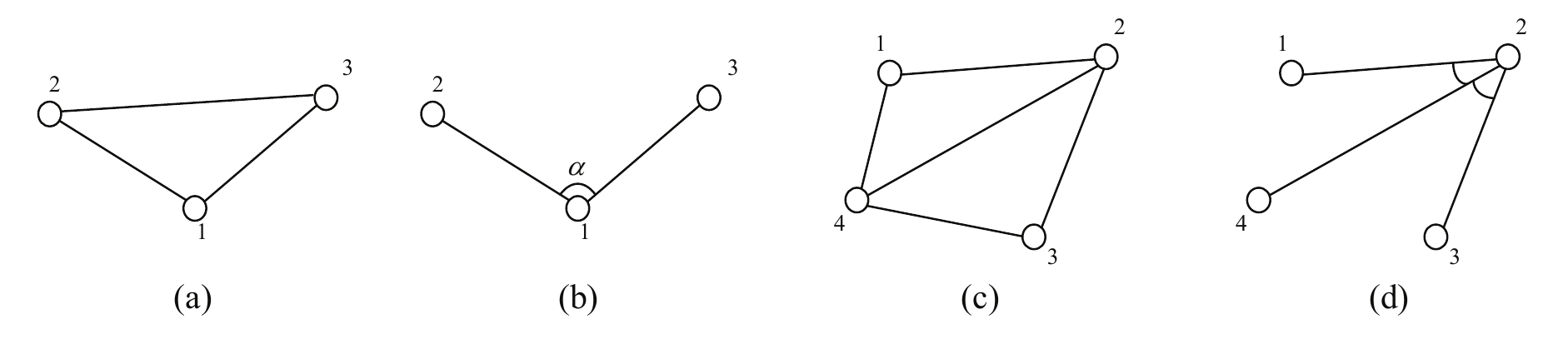}
\caption{Four frameworks in $\mathbb{R}^2$. (a) A minimally and globally rigid framework. (b) A non-rigid framework. (c)  A minimally rigid framework. (d) A non-rigid framework.}
\label{fig 4 examples}
\end{figure}

We look at a simple example first. Fig. \ref{fig 4 examples} presents several frameworks embedded in the plane. Observe that the framework in Fig. 1(a) is minimally and globally rigid. One can see that if the information of edge $(2,3)$ is absent, as shown in Fig. 1(b), the resulting framework is non-rigid. Nevertheless, the geometric shape can still be uniquely determined up to rigid transformations by $||e_{12}||$, $||e_{13}||$ and $e^T_{12}e_{13}$.   This is due to the fact that $||e_{23}||^2=||-e_{12}+e_{13}||^2=||e_{12}||^2+||e_{13}||^2-2e_{12}^Te_{13}$. In fact, once $||e_{12}||$ and $||e_{13}||$ are both available, $e_{12}^Te_{13}$ is actually a constraint for the angle $\alpha$ subtended at vertex $i$. Similarly, the non-rigid framework with constraint set $\{||e_{21}||^2, e_{21}^Te_{23}, ||e_{23}||^2, e_{23}^Te_{24}, ||e_{24}||^2\}$ in Fig. 1(d) is sufficient to determine the minimally rigid framework in Fig. 1(c).

\subsection{Definitions associated with weak rigidity}\label{subsec:definitions}
Now we introduce a generalized version of rigidity by utilizing a different rigidity function. Here $e_{ij}^Te_{ik}$ is employed as a component of the modified rigidity function. Let $\mathcal{T}_\mathcal{G}=\{(i,j,k)\in\mathcal{V}^3: (i,j),(i,k)\in\mathcal{E}, j\leq k\}$. In many cases, it is sufficient to recognize a framework when the information of $e_{ij}^Te_{ik}$ for partial $(i,j,k)\in\mathcal{T}_\mathcal{G}$ is available. We use $\mathcal{T}_\mathcal{G}^*$ with $|\mathcal{T}_\mathcal{G}^*|=s$ to denote a subset of $\mathcal{T}_\mathcal{G}$ such that for each triple $(i,j,k)\in\mathcal{T}_\mathcal{G}^*$, $e_{ij}^Te_{ik}$ is a component of the modified rigidity function. The modified rigidity function $r_{\mathcal{G}}(\cdot):\mathbb{R}^{nd}\rightarrow\mathbb{R}^s$ is given by
\begin{equation}\label{rigidfun}
r_{\mathcal{G}}(p)=(\cdots,e_{ij}^Te_{ik},\cdots)^T, ~~(i,j,k)\in\mathcal{T}_\mathcal{G}^*.
\end{equation}
Note that for a framework $(\mathcal{G},p)$, the choice of $\mathcal{T}_\mathcal{G}^*$ is not unique. Moreover, whether $(\mathcal{G},p)$ can be determined by $(\ref{rigidfun})$ is directly dependent on $\mathcal{T}_\mathcal{G}^*$. Next, we give several definitions associated with weak rigidity.

\begin{definition}
$(\mathcal{G},p)$ and $(\mathcal{G},q)$ are weakly equivalent for a given $\mathcal{T}_\mathcal{G}^*$ if $(p_i-p_j)^T(p_i-p_k)=(q_i-q_j)^T(q_i-q_k)$ for all $(i,j,k)\in\mathcal{T}_\mathcal{G}^*$.
\end{definition}
\begin{definition}\label{def:weakly congruent}
$(\mathcal{G},p)$ and $(\mathcal{G},q)$ are weakly congruent if $(p_i-p_j)^T(p_i-p_k)=(q_i-q_j)^T(q_i-q_k)$ for all $i,j,k\in\mathcal{V}$.
\end{definition}
\begin{definition}\label{def:weak rigidity}
A framework $(\mathcal{G},p)$ is weakly rigid if there exists a neighborhood $U_p$ of $p$ such that for any $q\in U_p$ and some $\mathcal{T}_\mathcal{G}^*$, if $(\mathcal{G},p)$ is weakly equivalent to $(\mathcal{G},q)$, then they are weakly congruent.
\end{definition}
\begin{definition}
A framework $(\mathcal{G},p)$ is globally weakly rigid if for any $q\in\mathbb{R}^{nd}$ and some $\mathcal{T}_\mathcal{G}^*$, once $(\mathcal{G},p)$ is weakly equivalent to $(\mathcal{G},q)$, they are weakly congruent.
\end{definition}
\begin{definition}
A framework $(\mathcal{G},p)$ is minimally weakly rigid if $(\mathcal{G},p)$ is weakly rigid, and deletion of any edge will make $(\mathcal{G},p)$ not weakly rigid.
\end{definition}

By these definitions, the framework in Fig. \ref{fig 4 examples} (b) with $\mathcal{T}_\mathcal{G}^*=\{(1,2,2),(1,3,3),$ $(1,2,3)\}$ is globally and minimally weakly rigid, while the framework in Fig. \ref{fig 4 examples} (d) with $\mathcal{T}_\mathcal{G}^*=\{(2,1,1),$ $(2,3,3),(2,4,4),(2,1,4),(2,3,4)\}$ is minimally weakly rigid. Note that these two frameworks are both non-rigid.

In \cite{Stacey16}, the authors defined a concept of ``generalized rigidity" for a multi-agent formation by generalizing both the state space of each agent and the relative state constraints characterizing the formation. Since the constraint $(p_i-p_j)^T(p_i-p_k)$ can be viewed as a specific form of a function of all vertices' coordinates, weak rigidity is a special case of ``generalized rigidity". The concept of ``weak rigidity'' is also defined in \cite{Park17} by adding angle constraints into the distance-based rigidity function.  Given a graph $\mathcal{G} =(\mathcal{V}, \mathcal{E}) $, an augmented graph $\bar{\mathcal{G}}=(\mathcal{V},\bar{\mathcal{E}})$ is constructed, in which $\bar{\mathcal{E}}$ is obtained by adding the edge $(j,k)$ into the edge set $\mathcal{E}$ if the angle between $p_i-p_j$ and $p_i-p_k$ is used as an entry of the weak rigidity function. Then the relationship between weak rigidity of $(\mathcal{G},p)$ and rigidity of $(\bar{\mathcal{G}},p)$ is studied. Note that in \cite{Park17}, $\| p_i-p_j \|$ is always available for any edge $(i,j)$. When $\| p_i - p_j \|$ and $\| p_i - p_k \|$ are both known, $(p_i-p_j)^T(p_i-p_k)$ is actually a constraint on the angle $\angle p^i_{jk}$ between $p_i-p_j$ and $p_i-p_k$. Therefore, their definition can be viewed as a special case of our definition. In our paper, the relationship between weak rigidity and rigidity is also discussed, but is given for the same framework, not by introducing an augmented graph. See Theorem \ref{rigidtogrigid} and its proof in Subsection 3.3. Actually, our work focuses on exploring properties of weak rigidity in depth. We show that compared to rigidity, weak rigidity defined in our paper has  nice properties and is easier to check, see Subsections 3.2-3.5. Moreover, the applications of the proposed weak rigidity theory on formation stabilization are studied, see Section 4.

To preserve the invariance of $r_{\mathcal{G}}(p)$, an infinitesimal motion $v=(v_1^T,\cdots,v_n^T)^T\in\mathbb{R}^{nd}$ should satisfy
\begin{equation}\label{ginfinitesimal}
(v_i-v_j)^Te_{ik}+e_{ij}^T(v_i-v_k)=0,~~(i,j,k)\in\mathcal{T}_\mathcal{G}^*.
\end{equation}

Equation (\ref{ginfinitesimal}) can be equivalently written as $\dot{r}_{\mathcal{G}}=\frac{\partial r_{\mathcal{G}}}{\partial p}\dot{p}= R_w(p)\dot{p}=0$, where $R_w(p)\triangleq\frac{\partial r_{\mathcal{G}}}{\partial p}\in\mathbb{R}^{s\times nd}$ is called the weak rigidity matrix. Let $g_{\mathcal{K}}$ be the distance rigidity function corresponding to the complete graph $\mathcal{K}$, since it always holds $e_{ij}^Te_{ik}=(||e_{ij}||^2+||e_{ik}||^2-||e_{jk}||^2)/2$, there exists a constant matrix $M\in\mathbb{R}^{|\mathcal{T}_{\mathcal{G}}^*|\times (n(n-1)/2)}$ such that $r_{\mathcal{G}}=Mg_{\mathcal{K}}$. Note that a distance rigidity function in $\mathbb{R}^d$ is $\textrm{\textrm{SE(d)}}$ invariant, i.e., invariant under translations and rotations. As a result, $r_{\mathcal{G}}$ is also $\textrm{SE(d)}$ invariant. It is natural to obtain that the trivial motion space for weak rigidity, which consists of infinitesimal motions such that (\ref{ginfinitesimal}) always holds, is identical to the one for rigidity. We then have the following lemma directly.

\begin{lemma}\label{le null(R_w)}
	The trivial motion space for weak rigidity is $\mathcal{S}=\mathcal{S}_r\cup\mathcal{S}_t$, where $\mathcal{S}_r=\{(I_n\otimes A)p: A+A^T=0, A\in\mathbb{R}^{d\times d}\}$ is the space including all infinitesimal motions that correspond to rotational motions, and $\mathcal{S}_t=\{\mathbf{1}_n\otimes q_i: q_i=(0,\cdots,0,1(i\textmd{th}),0,\cdots,0)^T\in\mathbb{R}^d, i=1,\ldots,d\}$ is the space including all infinitesimal motions that correspond to translational motions.
\end{lemma}

The specific forms of rotation motion space and translation motion space have also been given in \cite{Zelazo15} and \cite{Sun16auto} for the case when $d=2,3$. Note that the rotational motion mentioned in Lemma \ref{le null(R_w)} is actually an infinitesimal rotation, also the translational motion denotes an infinitesimal translation. It is easy to see that the trivial motion space $\mathcal{S}$ always belongs to $\textrm{null}(R_w)$, thus $\textrm{rank}(R_w)\leq nd-d(d+1)/2$. We present the following definition for infinitesimal weak rigidity.

\begin{definition}\label{de infinitesimal}
A framework $(\mathcal{G},p)$ is infinitesimally weakly rigid if there exists a $\mathcal{T}_\mathcal{G}^*$ such that every infinitesimal motion satisfying (\ref{ginfinitesimal}) is trivial, or equivalently, $\textrm{rank}(R_w)=nd-d(d+1)/2$.
\end{definition}

Observe that if each component of $r_{\mathcal{G}}$ in (\ref{rigidfun}) is a length constraint of an edge, or equivalently if $j=k$ for all $(i,j,k)\in\mathcal{T}_\mathcal{G}^*$, then the weak rigidity function becomes a rigidity function. Similar to infinitesimal rigidity, an implicit condition for infinitesimal weak rigidity of $(\mathcal{G},p)$ is $s=|\mathcal{T}_\mathcal{G}^*|\geq nd-d(d+1)/2$. That is, compared to the distance-based rigidity, in characterizing a framework without nontrivial infinitesimal motions, the number of edges involved in the weak rigidity function can be reduced, but the number of entries in the weak rigidity function cannot be reduced.


Suppose $\mathcal{G}$ has a spanning tree $T_r=(\mathcal{V},\mathcal{E}_{tr})$. Let $e_{tr}=(\cdots,e_{ij}^T,\cdots)^T$, $(i,j)\in\mathcal{E}_{tr}$, $r_{T_r}=(\cdots,e_{ij}^Te_{ik},\cdots)^T$, $(i,j),(i,k)\in\mathcal{E}_{tr}$, $(i,j,k)\in\mathcal{T}_\mathcal{G}^*$.  Define
\begin{equation} \label{ew}
R^{tr}_e(p) \triangleq \frac{\partial r_{T_r}}{\partial e_{tr}},  \  \  \  R_w^{tr}(p) \triangleq \frac{\partial r_{T_r}}{\partial p}.
\end{equation} 
By the chain rule we have
\begin{equation}\label{R_w_e_H}
R_w^{tr}(p)=\frac{\partial r_{T_r}}{\partial e_{tr}}\frac{\partial e_{tr}}{\partial p}= R^{tr}_e(p)\bar{H}_{tr},
\end{equation}
where $\bar{H}_{tr}\triangleq\frac{\partial e_{tr}}{\partial p}=H_{tr}\otimes I_d$ with $H_{tr}$ being the incidence matrix of $T_r$. It is easy to see that $R_w^{tr}$ is a submatrix of $R_w$. Hence $\textrm{rank}(R_w)\geq \textrm{rank}(R_w^{tr})$. That is, if $\textrm{rank}(R_w^{tr})=2n-d(d+1)/2$, infinitesimal weak rigidity can be guaranteed.

Now we try to fix $\textrm{rank}(R_w^{tr})$ by restricting $\textrm{rank}(R^{tr}_e)$.  Rewrite $e_{tr}$ as $e_{tr} =(\alpha_1^T,\cdots,\alpha_{n-1}^T)^T$, where for each $i\in\{1,\cdots,n-1\}$, $\alpha_i=e_{jk}$ for some $(j,k)\in\mathcal{E}_{tr}$. In fact, by regarding $\alpha_i^T\alpha_j$ as the distance of edges $\alpha_i$ and $\alpha_j$, see \cite{Singer10}, $R^{tr}_e\triangleq\frac{\partial r_{T_r}}{\partial e_{tr}}$ in (\ref{R_w_e_H}) can be viewed as a distance rigidity matrix corresponding to the following rigidity function
$$r_{T_r}=(\cdots,\alpha_i^T\alpha_j,\cdots)^T.$$
Different from Euclidean distance, $\alpha_i^T\alpha_j$ cannot be preserved during an identical translation of $\alpha_i$ and $\alpha_j$, i.e., $\alpha_i^T\alpha_j\neq (\alpha_i+c)^T(\alpha_j+c)$ for some $c\in\mathbb{R}^d$. Therefore, a trivial motion of $\alpha_i$ and $\alpha_j$ for preserving $\alpha_i^T\alpha_j$ can only be rotation. Since the dimension of the space spanned by independent rotations is $d(d-1)/2$, one has $\textrm{rank}(R^{tr}_e)\leq |\mathcal{E}_{tr}|d-d(d-1)/2=nd-d(d+1)/2$. In fact, if we restrict $\textrm{rank}(R^{tr}_e)=nd-d(d+1)/2$, then $\textrm{null}(R^{tr}_e)=\{\bar{H}_{tr}v\in\mathbb{R}^{(n-1)d}:v\in\mathcal{S}_r\}$, together with $\textrm{null}(\bar{H}_{tr})=\mathcal{S}_t$, we have $\textrm{null}(R_w^{tr})=\mathcal{S}$. Therefore, we have the following lemma.

\begin{lemma}\label{le rank(R_e)}
Given a framework $(\mathcal{G},p)$, if there exists a spanning tree $T_r=(\mathcal{V},\mathcal{E}_{tr})$ in graph $\mathcal{G}$ and a $\mathcal{T}_\mathcal{G}^*$ such that $\textrm{rank}(R^{tr}_e)=nd-d(d+1)/2$, then $(\mathcal{G},p)$ is infinitesimally weakly rigid.
\end{lemma}

\begin{remark}\label{re condition on R_e}
The rank condition on $R^{tr}_e$ cannot be used to check infinitesimal rigidity since a framework embedded by a tree can never be rigid. However, it is efficient to determine infinitesimal weak rigidity in many circumstances. This is a critical difference between distance-based rigidity and weak rigidity, also showing that fewer edges are required for guaranteeing weak rigidity of a framework.
\end{remark}

\subsection{Construction of a minimal $\mathcal{T}_{\mathcal{G}}^*$ for infinitesimal weak rigidity in the plane}

In Subsection \ref{subsec:definitions}, we show that for a framework $(\mathcal{G},p)$, a subset $\mathcal{T}_{\mathcal{G}}^*$ of $\mathcal{T}_{\mathcal{G}}$ is often sufficient for the weak rigidity function to determine the weak rigidity of $(\mathcal{G},p)$, thus there often exist redundant elements in $\mathcal{T}_{\mathcal{G}}$. In this subsection, we will show how to construct a minimal $\mathcal{T}_{\mathcal{G}}^*$ (i.e., $\mathcal{T}_{\mathcal{G}}^*$ contains minimal number of elements) such that a planar framework $(\mathcal{G},p)$ with $\mathcal{T}_{\mathcal{G}}^*$ is infinitesimally weakly rigid. Before showing this, a necessary and sufficient graphical condition for infinitesimal weak rigidity in $\mathbb{R}^2$ is presented in the following theorem.

\begin{theorem}\label{th ns condition for GIS}
A framework $(\mathcal{G},p)$ with $n\geq 3$ vertices in $\mathbb{R}^2$ is infinitesimally weakly rigid if and only if $\mathcal{G}$ is connected, and for any $i\in\mathcal{V}$ with $|\mathcal{N}_i|\geq2$, there exist at least two vertices $j,k\in\mathcal{N}_i$ such that $e_{ij}$ and $e_{ik}$ are not collinear.
\end{theorem}
\begin{proof} (Necessity) Actually, the necessity condition holds for frameworks in $\mathbb{R}^d$ with any $d\geq2$. Therefore we give a proof in the general case. We first show that $\mathcal{G}$ is connected. Suppose this is not true, then for any selected $\mathcal{T}_\mathcal{G}^*$, each independent connected subgraph can rotate independently while preserving $R_w(p)\dot{p}=0$. This conflicts with Definition \ref{de infinitesimal}. Therefore, graph $\mathcal{G}$ must be connected. Next we prove the second part. Suppose that vertex $i$ has more than two neighbors, and all $e_{ij}$, $j\in\mathcal{N}_i$ are collinear. Notice that $e_{ij}$ cannot all be zero; otherwise, $\textrm{rank}(R_w(p))<nd-d(d+1)/2$ and so the framework $(\mathcal{G},p)$ is not infinitesimally weakly rigid. Let $e_{ik}$ be a nonzero vector, then $e_{ij}=c_je_{ik}$ with a nonzero $c_j\in\mathbb{R}$ for all $j\in\mathcal{N}_i$. Let $A\in\mathbb{R}^{d\times d}$ be a nontrivial skew-symmetric matrix, we can obtain that $q=(\mathbf{0},\cdots,\mathbf{0},(Ae_{ik})^T,\mathbf{0},\cdots,\mathbf{0})^T\in \textrm{null}(R_w)$, where the components of $(Ae_{ik})^T$ are $(i-1)d+1$ to $id$ components of $q$. Due to Lemma \ref{le null(R_w)}, $q$ is not a trivial motion, a contradiction with infinitesimal weak rigidity of $(\mathcal{G},p)$ arises.

(Sufficiency) We first claim that there exists a spanning tree $T_r=(\mathcal{V},\mathcal{E}_{tr})$ of $\mathcal{G}$ satisfying that for any $i\in\mathcal{V}$, there are at least two vertices $j,k\in\mathcal{V}$ such that $(i,j),(i,k)\in\mathcal{E}_{tr}$, $e_{ij}$ and $e_{ik}$ are not collinear. Suppose this is not true. Then there exists a vertex $i$ with $|\mathcal{N}_i|\geq3$ in a cycle of $\mathcal{G}$, the deletion of any edge $(i,j)$ in this cycle will make $e_{il}$, $l\in\mathcal{N}_i\setminus \{j\}$ be all collinear. This implies that $e_{ij}$ and $e_{il}$ are not collinear for all $l\in\mathcal{N}_i\setminus \{j\}$. Note that there must be two edges involving $i$ in the cycle. Without loss of generality, let $(i,j),(i,k)$ be these two edges. Then one can see that deleting $(i,k)$ rather than $(i,j)$ can also eliminate the cycle and make the vectors $e_{il}$, $l\in\mathcal{N}_i\setminus \{k\}$ be not all collinear, which is a contradiction. Hence the existence of $T_r$ is proved. By Lemma \ref{le rank(R_e)}, it suffices to show $\textrm{rank}(R_e^{tr})=2n-d(d+1)/2$.

By virtue of the above conclusion, the sufficiency can be proved in the case when $\mathcal{G}$ is a tree and generality is not lost. Now we regard $\mathcal{G}$ as a tree. It is only required to construct a set $\mathcal{T}_\mathcal{G}^*$, such that $\textrm{rank}(R_e)=\frac{\partial r_{\mathcal{G}}}{\partial e}=|\mathcal{E}|d-d(d-1)/2=2m-1$, where $e=(\cdots,e^T_{ij},\cdots)^T$, $(i,j)\in\mathcal{E}$. Let $\mathcal{T}_\mathcal{G}^*=\{(i,j,j)\in\mathcal{V}^3:(i,j)\in\mathcal{E}\}\cup\mathcal{F}$, then $r_{\mathcal{G}}=(g_{\mathcal{G}}^T,\bar{r}^T)^T$, where $\bar{r}=(\cdots,e_{ij}^Te_{ik},\cdots)^T$, $(i,j,k)\in\mathcal{F}$. Finding elements of $\mathcal{F}$ is equivalent to finding components of $\bar{r}$. Next we present an approach to constructing $\bar{r}$.

Let $\mathcal{H}\subseteq\mathcal{V}$ be the set of internal vertices, i.e., the vertices with more than two neighbors in $\mathcal{G}$. That is, $|\mathcal{N}_i|\geq2$ for all $i\in\mathcal{H}$. Since $n\geq3$ and $\mathcal{G}$ is connected, $|\mathcal{H}|\neq\varnothing$.  Note that it always holds that $$2|\mathcal{E}|=\sum_{i\in\mathcal{V}}|\mathcal{N}_i|=|\mathcal{V}|-|\mathcal{H}|+\sum_{i\in\mathcal{H}}|\mathcal{N}_i|.$$ Since $\mathcal{G}$ is a tree, it holds that $n=|\mathcal{V}|=|\mathcal{E}|+1=m+1$. It follows that $$\sum_{i\in\mathcal{H}}|\mathcal{N}_i|-|\mathcal{H}|=m-1.$$ In fact, we can give $|\mathcal{N}_i|-1$ components of $\bar{r}$ for each $i\in\mathcal{H}$, which are pairwise inner products of relative position vectors corresponding to the $|\mathcal{N}_i|$ edges. Note that for $i\in\mathcal{H}$, we can always divide $\mathcal{N}_i$ into two disjoint sets $\hat{\mathcal{N}}_i$ and $\check{\mathcal{N}}_i$ such that $e_{ij}$ and $e_{ik}$ are not collinear for any $j\in\hat{\mathcal{N}}_i$, $k\in\check{\mathcal{N}}_i$.

We first select a vertex $j_i\in\hat{\mathcal{N}}_i$ randomly, let $e_{ij_i}^Te_{ik}$, $k\in\check{\mathcal{N}}_i$ be partial components of $\bar{r}$. Next we select a vertex $k_i\in\check{\mathcal{N}}_i$ randomly, let $e_{ij}^Te_{ik_i}$, $j\in\hat{\mathcal{N}}_i\setminus\{j_i\}$ be the components of $\bar{r}$. Then we have presented an approach for giving $|\hat{\mathcal{N}}_i|+|\check{\mathcal{N}}_i|-1=|\mathcal{N}_i|-1$ components of $\bar{r}$ for a vertex $i\in\mathcal{H}$. By this approach, we can totally give $\sum_{i\in\mathcal{H}}\mathcal{N}_i-|\mathcal{H}|=m-1$ components of $\bar{r}$. Now we prove that the rows of $R_e$ corresponding to these $m-1$ constraints, which are actually the rows of $\frac{\partial \bar{r}}{\partial e}$, together with the rows of $\frac{\partial g_\mathcal{G}}{\partial e}$, are linearly independent. Suppose the following holds for some scalars $l_{ij}$, $\bar{l}_{ij}$,
\begin{equation}\label{linear combination'}
\sum_{i\in\mathcal{H}}\big(\sum_{k\in\check{\mathcal{N}}_i}l_{ik}\frac{\partial e_{ij_i}^Te_{ik}}{\partial e}+\sum_{j\in\hat{\mathcal{N}}_i\setminus\{j_i\}}l_{ij}\frac{\partial e_{ij}^Te_{ik_i}}{\partial e}+\sum_{h\in\mathcal{N}_i}\bar{l}_{ih}\frac{\partial ||e_{ih}||^2}{\partial e}\big)=0.
\end{equation}
Let $\bar{\mathcal{H}}\subseteq\mathcal{H}$ be the set such that for any vertex $i\in\bar{\mathcal{H}}$, $\mathcal{N}_i$ includes at least one leaf vertex, where a leaf vertex is a vertex having only one neighbor. In fact, there must exist a leaf vertex $j\in\mathcal{N}_i$ for some $i\in\bar{\mathcal{H}}$, such that only one component of $\bar{r}$ involves $e_{ij}$. To show this, we claim that one of the following statements must be true.

(i) There exists a vertex $i\in\bar{\mathcal{H}}$ such that $|\mathcal{N}_i|=2$.

(ii) There exists a vertex $i\in\bar{\mathcal{H}}$, such that $|\mathcal{N}_i|=3$, $\mathcal{N}_i$ includes at least two leaf vertices.

(iii) There exists a vertex $i\in\bar{\mathcal{H}}$, such that $|\mathcal{N}_i|\geq4$, $\mathcal{N}_i$ includes at least three leaf vertices.

Suppose that all the above statements are not true. Then $\bar{\mathcal{H}}$ can be divided into three sets, i.e., $\bar{\mathcal{H}}=\bar{\mathcal{H}}_1\cup\bar{\mathcal{H}}_2\cup\bar{\mathcal{H}}_3$, such that if $i\in\bar{\mathcal{H}}_1$, then $|\mathcal{N}_i|=3$ and $\mathcal{N}_i$ includes one leaf vertex exactly; if $i\in\bar{\mathcal{H}}_2$, then $|\mathcal{N}_i|\geq4$ and $\mathcal{N}_i$ includes one leaf vertex exactly; if $i\in\bar{\mathcal{H}}_3$, then $|\mathcal{N}_i|\geq4$ and $\mathcal{N}_i$ includes two leaf vertices exactly. Let $n_i$ be the number of leaf vertices having a neighbor in $\bar{\mathcal{H}}_i$, $i=1,2,3$. Then we have $|\bar{\mathcal{H}}_1|=n_1$, $|\bar{\mathcal{H}}_2|=n_2$, $|\bar{\mathcal{H}}_3|=n_3/2$, and $n_1+n_2+n_3=n-|\mathcal{H}|$. It follows that
\begin{equation*}
\begin{split}
2m&=\sum_{i\in\mathcal{V}}|\mathcal{N}_i|\geq n-|\mathcal{H}|+3|\bar{\mathcal{H}}_1|+4|\bar{\mathcal{H}}_2|+4|\bar{\mathcal{H}}_3|+2(|\mathcal{H}|-|\bar{\mathcal{H}}|)\\
&=n+|\mathcal{H}|+n_1+2n_2+n_3\\
&\geq n+|\mathcal{H}|+n-|\mathcal{H}|=2n.
\end{split}
\end{equation*}
This conflicts with $m=n-1$. Hence, there is at least one true statement among (i), (ii) and (iii). Now we discuss in the following three cases.

Case 1. (i) holds. Let $\mathcal{N}_i=\{j,k\}$, where $j$ is a leaf vertex. Then it is obvious that there is only one component $e_{ij}^Te_{ik}$ in $\bar{r}$ involving $j$.

Case 2. (ii) holds. There are $|\mathcal{N}_i|-1=2$ components selected from $e_{ij}^Te_{ik}$. Since $\mathcal{N}_i$ includes two leaf vertices, there must exist one leaf vertex $l\in\mathcal{N}_i$ such that $e_{il}$ is involved by only one of the two components.

Case 3. (iii) holds. Note that for any $i\in\mathcal{H}$, only two vertices $j_i,k_i\in\mathcal{N}_i$ are possibly involved by more than two components selected from $e_{ij}^Te_{ik}$, $j,k\in\mathcal{N}_i$. Hence, there must exist at least one leaf vertex $l\in\mathcal{N}_i$ such that $e_{il}$ is involved by only one component of $\bar{r}$.

So far we have proved that there always exists a leaf vertex $j\in\mathcal{N}_i$ for some $i\in\bar{\mathcal{H}}$, such that only one component $e_{ij}^Te_{ik}$ of $\bar{r}$ involves $e_{ij}$. Observe that there are only two nonzero rows in $\frac{\partial r_{\mathcal{G}}}{\partial e_{ij}}$, i.e., $e_{ij}^T$ and $e_{ik}^T$. Since $e_{ij}$ and $e_{ik}$ are not collinear, the validity of (\ref{linear combination'}) implies $l_{ij}=\bar{l}_{ij}=0$. Note that after deleting vertex $j$ and edge $(i,j)$, $\mathcal{G}'=(\mathcal{V}\setminus\{j\},\mathcal{E}\setminus\{(i,j)\})$ is another tree. By the aforementioned approach, we can prove that once (\ref{linear combination'}) holds, there is a leaf vertex $j'\in\mathcal{V}\setminus\{j\}$, such that $l_{i'j'}=\bar{l}_{i'j'}=0$. By repeating this process, we can finally obtain that $l_{ij}=\bar{l}_{ij}=0$ for all $(i,j)\in\mathcal{E}$. As a result, $\textrm{rank}(R_e)=m+m-1=2m-1$. That is, $(\mathcal{G},p)$ is infinitesimally weakly rigid.
\end{proof}

\begin{figure}
\centering
\includegraphics[width=13cm]{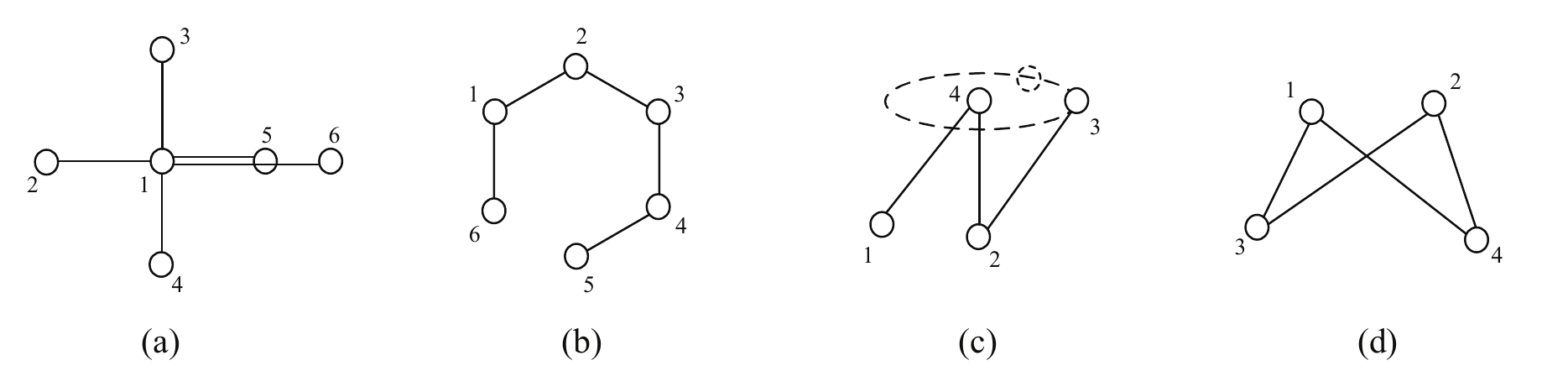}
\caption{Four non-rigid frameworks. (a) A minimally infinitesimally weakly rigid framework in $\mathbb{R}^2$ with a minimal $\mathcal{T}_\mathcal{G}^*=\{(1,2,3),(1,2,4),(1,3,5),(1,3,6),(1,j,j), j\in\mathcal{N}_1\}$. (b) A minimally infinitesimally weakly rigid framework in $\mathbb{R}^2$ with $\mathcal{T}_\mathcal{G}^*=\{(1,2,6),(2,1,3),(3,2,4),(4,3,5),(i,j,j),(i,j)\in\mathcal{E},i>j\}$. (c) A framework which is not weakly rigid in $\mathbb{R}^3$. (d) A minimally and globally weakly rigid framework in $\mathbb{R}^3$ with $\mathcal{T}_\mathcal{G}^*=\{(1,3,3),(1,4,4),(2,3,3),(2,4,4),(1,3,4),(3,1,2)\}$.}
\label{fig conditionforR_e}
\end{figure}

Two infinitesimally weakly rigid frameworks are given in Fig. \ref{fig conditionforR_e} (a) and (b) to demonstrate Theorem \ref{th ns condition for GIS}.

It is important to note that by virtue of Theorem \ref{th ns condition for GIS}, the sufficiency condition in Lemma \ref{le rank(R_e)} for infinitesimal weak rigidity is also necessary when $d=2$. One may ask whether the sufficiency of Theorem \ref{th ns condition for GIS} and necessity of Lemma \ref{le rank(R_e)} hold for frameworks in $\mathbb{R}^d$ with $d\geq3$. The answer is not. We show two counter-examples as follows.

\textbf{Two Counter-examples:} In Fig. \ref{fig conditionforR_e}, two frameworks in $\mathbb{R}^3$ are shown in (c) and (d). In (c),  for vertex $4$, $e_{41}$ and $e_{42}$ are not collinear. Similarly, for vertex $2$, $e_{23}$ and $e_{24}$ are not collinear.  However, vertex $3$ can move along the dotted circle continuously while preserving the invariance of $r_{\mathcal{G}}=(||e_{14}||^2,||e_{24}||^2,||e_{23}||^2,$ $e_{41}^Te_{42},e_{23}^Te_{24})^T$, which implies that $\textrm{null}(R_w)$ includes nontrivial infinitesimal motions. Hence, the sufficiency of Theorem \ref{th ns condition for GIS} is invalid. The framework $(\mathcal{G},p)$ in (d) is infinitesimally weakly rigid. However, each spanning tree $T_r$ of $\mathcal{G}$ is isomorphic to the graph in (c), thus the framework $(T_r, p)$ also has nontrivial infinitesimal motions in $\mathbb{R}^3$. This implies that the necessity of Lemma \ref{le rank(R_e)} does not hold in $\mathbb{R}^3$.

\begin{remark} \label{rem:infinitesimal}
Observe that the necessary and sufficient condition in Theorem \ref{th ns condition for GIS} can be easily verified instead of examining the rank of a matrix, and is necessary but not sufficient for a framework to be infinitesimally rigid. This implies that infinitesimal weak rigidity is milder than infinitesimal rigidity for a framework. Moreover, the proof of Theorem \ref{th ns condition for GIS} actually provides an idea to constructing a minimal set $\mathcal{T}_\mathcal{G}^*$ for infinitesimal weak rigidity of $(\mathcal{G},p)$, thus will be used later.
\end{remark}

By Theorem \ref{th ns condition for GIS}, it is easy to design algorithms for examining infinitesimal weak rigidity of a planar framework. In the following, we will mainly focus on how to construct a minimal $\mathcal{T}_{\mathcal{G}}^*$ for a given infinitesimally weakly rigid framework $(\mathcal{G},p)$. According to Theorem \ref{th ns condition for GIS}, there must exist a spanning tree $T_r$ of $\mathcal{G}$ such that $(T_r,p)$ is minimally infinitesimally weakly rigid. We present Algorithm \ref{alg:spanning tree} to find $(T_r,p)$.

\begin{algorithm}
	\renewcommand{\algorithmicrequire}{\textbf{Input:}}
	\renewcommand{\algorithmicensure}{\textbf{Output:}}
	\caption{Finding a Minimally Infinitesimally Weakly Rigid Subframework}
	\label{alg:spanning tree}
	\begin{algorithmic}[1]
		\REQUIRE $\mathcal{G}=(\mathcal{V},\mathcal{E})$, $p=(p_1^T,\cdots,p_n^T)^T\in\mathbb{R}^{2n}$.
		\ENSURE $(T_r,p)$
		\STATE Initialize $\mathcal{V}_{tr}\gets\{a,b\}$, $\mathcal{E}_{tr}\gets(a,b)$, where $a$, $b$ are selected randomly such that $(a,b)\in\mathcal{E}$
		\WHILE{$|\mathcal{V}_{tr}|<n$}
		\STATE Select an edge $(i,j)\in\mathcal{E}$ such that $i\in\mathcal{V}_{tr}$, $j\in\mathcal{V}\setminus\mathcal{V}_{tr}$, there exists at least one edge $(i,k)\in\mathcal{E}_{tr}$ such that $p_i-p_j$ is not collinear with $p_i-p_k$
		\STATE $\mathcal{V}_{tr}\gets \{j\}$, $\mathcal{E}_{tr}\gets\{(i,j)\}$
		\ENDWHILE
		\STATE $T_r\gets(\mathcal{V}_{tr},\mathcal{E}_{tr})$
		\STATE \textbf{return} $(T_r,p)$
	\end{algorithmic}
\end{algorithm}

Notice that even for a minimally infinitesimally weakly rigid framework $(\mathcal{G},p)$, it may be possible that $|\mathcal{T}_\mathcal{G}|>nd-d(d+1)/2$. For example, in Fig. \ref{fig conditionforR_e} (a), a minimal $\mathcal{T}_\mathcal{G}^*$ should have $2n-3=9$ elements, but $|\mathcal{T}_\mathcal{G}|=15$.  In this case, we still have to choose suitable elements from $\mathcal{T}_\mathcal{G}$ to form a minimal $\mathcal{T}_\mathcal{G}^*$.  We adopt $\mathcal{T}_{\mathcal{G}}^\dagger$ to denote the minimal $\mathcal{T}_{T_r}^*$ for infinitesimal weak rigidity of $(T_r,p)$ generated by Algorithm \ref{alg:spanning tree}, which is also the minimal $\mathcal{T}_{\mathcal{G}}^*$ for infinitesimal weak rigidity of $(\mathcal{G},p)$. Algorithm \ref{alg: construction of T_G^*} is designed to construct $\mathcal{T}_{\mathcal{G}}^\dagger$.

\begin{algorithm}[htbp]
	\renewcommand{\algorithmicrequire}{\textbf{Input:}}
	\renewcommand{\algorithmicensure}{\textbf{Output:}}
	\caption{Construction of $\mathcal{T}_{\mathcal{G}}^\dagger$}
	\label{alg: construction of T_G^*}
	\begin{algorithmic}[1]
		\REQUIRE $T_r=(\mathcal{V}_{tr},\mathcal{E}_{tr})$, $p=(p_1^T,\cdots,p_n^T)^T\in\mathbb{R}^{2n}$.
		\ENSURE $\mathcal{T}_{\mathcal{G}}^\dagger$
		\STATE Initialize  $\mathcal{T}_{\mathcal{G}}^\dagger\gets\{(i,j,j)\in\mathcal{V}^3_{tr}:(i,j)\in\mathcal{E}_{tr}\}$
		\FORALL{$i\in\mathcal{V}_{tr}$}
		\STATE Compute the neighbor set of $i$ in $T_r$, i.e., $\mathcal{N}_i$. Proceed only if {$|\mathcal{N}_i|\geq2$}
		\STATE $j_i\gets\min\mathcal{N}_i$, $\hat{\mathcal{N}}_i\gets\{j_i\}\cup\{k\in\mathcal{N}_i:p_i-p_{j_i}~is~collinear~with~p_i-p_k\}$, $\check{\mathcal{N}}_i\gets\mathcal{N}_i\setminus\hat{\mathcal{N}}_i$. Proceed only if {$\check{\mathcal{N}}_i\neq\varnothing$}
		\STATE $\mathcal{T}_{\mathcal{G}}^\dagger\gets\mathcal{T}_{\mathcal{G}}^\dagger\cup (i,j_i,k)$ for all $k\in\check{\mathcal{N}}_i$. Proceed only if {$|\hat{\mathcal{N}}_i|>1$}
		\STATE Select $k_i$ from $\check{\mathcal{N}}_i$ randomly
		\FORALL{$j\in\hat{\mathcal{N}}_i\setminus\{j_i\}$}
		\STATE $\mathcal{T}_{\mathcal{G}}^\dagger\gets\mathcal{T}_{\mathcal{G}}^\dagger\cup (i,j,k_i)$ if $j<k_i$, $\mathcal{T}_{\mathcal{G}}^\dagger\gets\mathcal{T}_{\mathcal{G}}^\dagger\cup (i,k_i,j)$ otherwise
		\ENDFOR
		\ENDFOR
		\STATE \textbf{return} $\mathcal{T}_{\mathcal{G}}^\dagger$
	\end{algorithmic}
\end{algorithm}

By Theorem \ref{th ns condition for GIS}, it is easy to see that $\mathcal{T}_{\mathcal{G}}^\dagger$ generated by Algorithm \ref{alg: construction of T_G^*} is sufficient for determining infinitesimal weak rigidity of $(\mathcal{G},p)$. Moreover, $\mathcal{T}_{\mathcal{G}}^\dagger$ contains $2n-3$ elements exactly, thus is minimal for infinitesimal weak rigidity of $(\mathcal{G},p)$. In Fig. \ref{fig conditionforR_e} (a), the framework is minimally infinitesimally weakly rigid, thus $\mathcal{T}_{\mathcal{G}}^\dagger$ can be obtained by Algorithm \ref{alg: construction of T_G^*} directly, a possible $\mathcal{T}_{\mathcal{G}}^\dagger$ generated by Algorithm \ref{alg: construction of T_G^*} is shown in the caption. In Fig. \ref{fig conditionforR_e} (b), $|\mathcal{T}_\mathcal{G}|=9=2n-3$, therefore, $\mathcal{T}_{\mathcal{G}}^\dagger=\mathcal{T}_{\mathcal{G}}$.

\subsection{Comparisons between rigidity and weak rigidity}

Compared to dist-ance-based rigidity, the advantage of weak rigidity is that fewer edges are required to determine a shape. The following theorem shows that rigidity is sufficient for weak rigidity.

\begin{theorem}\label{rigidtogrigid}
	If a framework $(\mathcal{G},p)$ is (infinitesimally, globally, minimally) rigid, then it is (infinitesimally, globally, minimally) weakly rigid.
\end{theorem}
\begin{proof} We choose a $\mathcal{T}_\mathcal{G}^*$ such that $(i,j,j)\in\mathcal{T}_\mathcal{G}^*$ for all $(i,j)\in\mathcal{E}$, then it is obvious that $rank(R_w)\geq rank(R)$. Therefore, infinitesimal rigidity leads to infinitesimal weak rigidity. Note that the components of $r_{\mathcal{K}}$ can always be denoted by a linear combination of distance constraints, i.e., $e_{ij}^Te_{ik}=(||e_{ij}||^2+||e_{ik}||^2-||e_{jk}||^2)/2$, therefore, congruence implies weak congruence.
	
Suppose $(\mathcal{G},p)$ is rigid, and $(\mathcal{G},q)$ is an arbitrary framework which is weakly equivalent to $(\mathcal{G},p)$ for the above-mentioned $\mathcal{T}_\mathcal{G}^*$. Then they are also equivalent. From rigidity of $(\mathcal{G},p)$, there exits a neighborhood $U_p$ of $p$ such that for any $q\in U_p$, $(\mathcal{G},p)$ and $(\mathcal{G},q)$ are congruent, which in turn implies that $(\mathcal{G},p)$ and $(\mathcal{G},q)$ are weakly congruent. Therefore $(\mathcal{G},p)$ is weakly rigid. Similarly, one can obtain that global rigidity implies global weak rigidity, and minimal rigidity implies minimal weak rigidity.
\end{proof}

The converse of Theorem \ref{rigidtogrigid} is not true, which has been shown in Remarks \ref{re condition on R_e} and \ref{rem:infinitesimal}. This also implies that weak rigidity requires fewer edges in the graph than rigidity does.

\subsection{The connection between weak rigidity and rigidity: a matrix completion perspective}
Using graph rigidity theory with the rigidity function (\ref{origidfun}), a graph realization problem is actually equivalent to a completion problem of a Euclidean distance matrix (EDM) completion problem, see \cite{Laurent97,Singer10,Liberti14}. A matrix completion problem asks whether the unspecified entries of partially defined matrix can be completed to obtain a fully defined matrix satisfying a desired property. An EDM is a matrix whose entries are  the pairwise squared Euclidean distances among a set of $n$ points in $d-$dimensional space \cite{Laurent97}. For a framework $(\mathcal{G},p)$, we denote the corresponding EDM by $D(p)\in\mathbb{R}^{n\times n}$. It is easy to see that two frameworks $(\mathcal{G},p)$ and $(\mathcal{G},q)$ are congruent if and only if $D(p)=D(q)$. Therefore, a framework can be determined up to rigid transformations if and only if the corresponding EDM can be uniquely completed. The following theorem shows a relationship between weak congruence and congruence.

\begin{theorem}\label{th gconD}
Two frameworks $(\mathcal{G},p)$ and $(\mathcal{G},q)$ are weakly congruent if and only if they are congruent (i.e., $D(p)=D(q)$).
\end{theorem}
\begin{proof} The necessity is obvious from Definition \ref{def:weakly congruent}. The sufficiency has been proved in the proof of Theorem \ref{rigidtogrigid}.
\end{proof}

It is straightforward to obtain the following corollary.
\begin{corollary}\label{co gk=rk}
Given a framework $(\mathcal{K},p)$ in $\mathbb{R}^d$, it holds that $g_{\mathcal{K}}^{-1}(g_{\mathcal{K}}(p))=r_{\mathcal{K}}^{-1}(r_{\mathcal{K}}(p))$ for $\mathcal{T}_\mathcal{K}^*=\mathcal{T}_\mathcal{K}$.
\end{corollary}
\begin{proof} For any $q\in\mathbb{R}^{nd}$, it follows from Theorem \ref{th gconD} that $$q\in g_{\mathcal{K}}^{-1}(g_{\mathcal{K}}(p))\Leftrightarrow g_{\mathcal{K}}(p)=g_{\mathcal{K}}(q)\Leftrightarrow r_{\mathcal{K}}(p)=r_{\mathcal{K}}(q)\Leftrightarrow q\in r_{\mathcal{K}}^{-1}(r_{\mathcal{K}}(p)).$$
The proof is completed.
\end{proof}

Corollary \ref{co gk=rk} implies that given a globally weakly rigid framework $(\mathcal{G},p^*)$ and a globally rigid framework $(\bar{\mathcal{G}},p^*)$, although $\mathcal{G}$ may have fewer edges than $\bar{\mathcal{G}}$, it holds that $r_{\mathcal{G}}(p^*)$ and $g_{\bar{\mathcal{G}}}(p^*)$ determine an identical geometric shape up to translations, rotations, and reflections.

In fact, when we use the weak rigidity function (\ref{rigidfun}) to recognize frameworks in $\mathbb{R}^d$, the graph realization problem can be transformed to a positive semi-definite (PSD) matrix completion problem \cite{Laurent97}. More precisely, let $E(p)\in\mathbb{R}^{d\times m}$ be the corresponding matrix with each column being a relative location vector, i.e., $E=(\cdots,e_{ij},\cdots)\in\mathbb{R}^{d\times m}$. We can observe that each component of $r_{\mathcal{G}}(\cdot)$ is actually an entry of the gram matrix $\mathbf{E}=E^TE$. If we regard $D(e_{ij},e_{kl})=e_{ij}^Te_{kl}$ as the distance between $e_{ij}$ and $e_{kl}$, then $\mathbf{E}$ becomes the distance matrix to be completed. The following theorem shows that for a connected graph $\mathcal{G}$, the framework $(\mathcal{G},p)$ can be determined up to rigid transformations if and only if the gram matrix $\mathbf{E}(p)\in\mathbb{R}^{m\times m}$ can be uniquely completed.

\begin{theorem}\label{th D=E}
Given two frameworks $(\mathcal{G},p)$ and $(\mathcal{G},q)$, if $\mathcal{G}$ is connected, then $D(p)=D(q)$ if and only if $\mathbf{E}(p)=\mathbf{E}(q)$.
\end{theorem}

\begin{proof} (Necessity) Suppose that $||p_i-p_j||=||q_i-q_j||$ for any $i,j\in\mathcal{V}$. Due to the fact that $e_{ij}^Te_{kl}=\frac12(||e_{jk}||^2+||e_{il}||^2-||e_{ik}||^2-||e_{jl}||^2)$, we have $(p_i-p_j)^T(p_k-p_l)=(q_i-q_j)^T(q_k-q_l)$ for any $(i,j),(k,l)\in\mathcal{E}$.

(Sufficiency) Since $\mathcal{G}$ is connected, for any $i,k\in\mathcal{V}$, there exists a path $\mathcal{P}=\{(i,i_1), ~\cdots, ~(i_r,k)\}\subseteq\mathcal{E}$ . It follows that
\begin{equation*}\begin{split}
||p_i-p_k||^2&=||p_i-p_{i_1}+\cdots+p_{i_r}-p_{k}||^2\\
&=\sum_{(j,l),(u,v)\in\mathcal{P}}(p_j-p_l)^T(p_u-p_v)\\
&=\sum_{(j,l),(u,v)\in\mathcal{P}}(q_j-q_l)^T(q_u-q_v)\\
&=||q_i-q_k||^2.
\end{split}\end{equation*}
\end{proof}

Now we are ready to present two necessity conditions for infinitesimal weak rigidity as follows.
\begin{theorem}\label{th girtogr}
	If $(\mathcal{G},p)$ with $n\geq d+1$ is infinitesimally weakly rigid for $\mathcal{T}_\mathcal{G}^*$, then
	
	(i) $p_1,\cdots,p_n$ do not lie in a hyperplane of $\mathbb{R}^d$;
	
	(ii) $(\mathcal{G},p)$ is weakly rigid for $\mathcal{T}_\mathcal{G}^*$.
\end{theorem}

\begin{proof} (i) Suppose this is not true. Then there always exists a nonzero normal vector $\eta\in \mathbb{R}^d$ and some constant $c\in\mathbb{R}$ such that $\eta^Tp_i=c$ for all $i\in\mathcal{V}$. Then $e_{ij}^T\eta=0$ for any $i,j\in\mathcal{V}$, which implies that $q=(\eta^T,0,\cdots,0)^T\in \textrm{null}(R_w)$. However, if $\dot{p}=q$, only vertex $1$ has a nonzero velocity, hence $q$ is obviously neither a rotational motion nor a translational motion. That is, $q$ does not belong to the trivial motion space described in Lemma \ref{le null(R_w)}, a contradiction arises.
	
(ii) For the differentiable map $r_{\mathcal{G}}(\cdot):\mathbb{R}^{nd}\rightarrow\mathbb{R}^s$, infinitesimal weak rigidity of $(\mathcal{G},p)$ implies that $\dim(\textrm{null}(R_w))$ reaches its minimum and $\frac{\partial r_\mathcal{G}}{\partial p}$ has a maximal rank at $p$. As a result, $p$ is a regular point of $r_{\mathcal{G}}$. \cite[Proposition 2]{Asimow78} shows that there exists a neighborhood $U_p$ of $p$, such that $r_\mathcal{G}^{-1}(r_{\mathcal{G}}(p))\cap U_p$ is a differentiable manifold of dimension $nd-\textrm{rank}(\frac{\partial r_{\mathcal{G}}}{\partial p})=d(d+1)/2$. Let $M=g_\mathcal{K}^{-1}(g_\mathcal{K}(p))$, where $\mathcal{K}$ is the complete graph with vertex set $\mathcal{V}$. The proof in \cite[Theorem]{Asimow78} shows that $M$ is a manifold of dimension $d(d+1)/2-(d-a)(d-a-1)/2=(a+1)(2d-a)/2$, where $a$ is the dimension of the affine hull of $\{p_1,\cdots,p_n\}$. According to Corollary \ref{co gk=rk}, it holds that $M=r_\mathcal{K}^{-1}(r_\mathcal{K}(p))$ with $\mathcal{T}_{\mathcal{G}}^*=\mathcal{T}_\mathcal{G}$. Note that $M\cap U_p$ is a submanifold of $r^{-1}_{\mathcal{G}}(r_{\mathcal{G}}(p))\cap U_p$ and they are equal if $a=d$ or $a=d-1$. The validity of (i) implies that $a=d$. Hence, $(\mathcal{G},p)$ is weakly rigid.
\end{proof}

Similar to traditional graph rigidity, weak rigidity cannot induce infinitesimal weak rigidity. A typical counter-example is a framework $(\mathcal{G},p)$ with $|\mathcal{V}|\geq d+1$ in $\mathbb{R}^d$, where $\mathcal{G}$ is a complete graph,  $e_{ij}$ for all $(i,j)\in\mathcal{E}$ lie on a hyperplane. In this case, $(\mathcal{G},p)$ is globally rigid and globally weakly rigid. However, when we let the normal vector to the hyperplane be the velocity of one vertex and zero be the velocity of all the other vertices, a nontrivial motion is constructed. Hence infinitesimal weak rigidity is not guaranteed.

\begin{remark}\label{re congruence}
By virtue of Theorems \ref{th gconD}, \ref{th D=E} and \ref{th girtogr}, once a framework $(\mathcal{G},p)$ is infinitesimally weakly rigid, there exists a neighborhood $U_p$ of $p\in\mathbb{R}^d$, such that if $q\in U_p$ and $r_{\mathcal{G}}(p)=r_{\mathcal{G}}(q)$, then $\mathbf{E}(p)=\mathbf{E}(q)$. Note that Theorem \ref{th girtogr} (i) implies $rank(E(p))=d$. Hence, the Cholesky decomposition of $\mathbf{E}(p)$ determines $E(p)$ uniquely up to an  orthogonal matrix $A\in\mathbb{R}^{d\times d}$. It follows that  $p_i-p_j=A(q_i-q_j)$ for all $(i,j)\in\mathcal{E}$. We then have $p_i=Aq_i+c$ for some $c\in\mathbb{R}^d$. If the determinant of $A$ is 1, then $A\in \textrm{SO(d)}$ is a rotation matrix, otherwise $A$ can be written as the product of a reflection matrix and a rotation matrix. The vector $c$ can be regarded as a translation vector. As a result, $(\mathcal{G},p)$ can be obtained by a rigid transformation from $(\mathcal{G},q)$.
\end{remark}

\subsection{Generic property}\label{subsec:generic property}

In \cite{Asimow78,Connelly05}, the authors show that rigidity is a generic property of the graph. In other words, for any graph $\mathcal{G}$, if $(\mathcal{G},p)$ is rigid for some generic configuration $p\in\mathbb{R}^{nd}$, then $(\mathcal{G},q)$ is rigid for any generic configuration $q\in\mathbb{R}^{nd}$. Here a configuration $p=(p_1^T,\cdots,p_n^T)\in\mathbb{R}^{nd}$ is said to be generic if its $nd$ coordinates are algebraically independent over integers \cite{Connelly05}. A vector $\alpha=(\alpha_1,\cdots,\alpha_{nd})$ is algebraically independent if there does not exist a nonzero polynomial $h(x_1,\cdots,x_{nd})$ with integer coefficients such that $h(\alpha_1,\cdots,\alpha_{nd})=0$. Since generic configurations form a dense subset of $\mathbb{R}^{nd}$, once $(\mathcal{G},p)$ is rigid for some generic configuration $p\in\mathbb{R}^{nd}$, $(\mathcal{G},q)$ is rigid for almost all configurations $q\in\mathbb{R}^{nd}$. In this subsection, we will show that for a framework $(\mathcal{G},p)$, both infinitesimal weak rigidity and weak rigidity are generic properties, thus are primarily determined by the graph $\mathcal{G}$, rather than the configuration $p$. Note that we only consider the case when $n\geq d+1$.

In analog to the discussions of generic rigidity  for graphs   in  \cite[Section 1.2]{Connelly05}, we present definitions of generic weak rigidity and generic infinitesimal weak rigidity for a graph as follows.
\begin{definition}
Graph $\mathcal{G}$ is said to be generically (infinitesimally) weakly rigid in $\mathbb{R}^d$ if for any generic configuration $p\in\mathbb{R}^{nd}$,  $(\mathcal{G},p)$ is (infinitesimally) weakly rigid.
\end{definition}

The following theorem shows that infinitesimal weak rigidity is a generic property of the graph.
\begin{theorem}\label{th ir is generic}
If $(\mathcal{G},p)$ is infinitesimally weakly rigid for a generic configuration $p\in\mathbb{R}^{nd}$,  then  graph $\mathcal{G}$ is generically infinitesimally weakly rigid in $\mathbb{R}^d$.
\end{theorem}
\begin{proof}
Let $p^*\in\mathbb{R}^{nd}$ be the generic configuration such that $(\mathcal{G},p^*)$ is infinitesimally weakly rigid for $\mathcal{T}_{\mathcal{G}}^*$. Then we have $\textrm{rank}(R_w(p^*))=\textrm{rank}(\frac{\partial r_{\mathcal{G}(p)}}{\partial p}|_{p=p^*})=nd-d(d+1)/2$. Since $\textrm{rank}(R_w(q))\leq nd-d(d+1)/2$ for all $q\in\mathbb{R}^{nd}$, we have $\max_{q\in\mathbb{R}^{nd}}$ $\textrm{rank}(R_w(q))=nd-d(d+1)/2$. Let $p\in\mathbb{R}^{nd}$ be a generic point distinct to $p^*$. The algebraic independence property of $p$ implies that each $(nd-d(d+1)/2)\times(nd-d(d+1)/2)$ minor of $R_w(p)$ cannot be zero. As a result, $p$ is a regular point, i.e., $\textrm{rank}(R_w(p))=\max_{q\in\mathbb{R}^{nd}}\textrm{rank}(R_w(q))=nd-d(d+1)/2$. Therefore, $(\mathcal{G},p)$ is infinitesimally weakly rigid. The proof is completed.
\end{proof}

To show that weak rigidity is also a generic property, we give the following result.

\begin{theorem}\label{th r with generic to ir}
If $(\mathcal{G},p)$ is weakly rigid for $\mathcal{T}_\mathcal{G}^*$ in $\mathbb{R}^d$, and $p\in\mathbb{R}^{nd}$ is generic, then $(\mathcal{G},p)$ is infinitesimally weakly rigid for $\mathcal{T}_\mathcal{G}^*$ in $\mathbb{R}^d$.
\end{theorem}
\begin{proof}
Let $\kappa=\max_{p\in\mathbb{R}^{nd}}\{\textrm{rank}(\frac{\partial r_{\mathcal{G}}(p)}{\partial p})\}$ with respect to $\mathcal{T}_{\mathcal{G}}^*$. From the proof of Theorem \ref{th ir is generic}, a generic configuration is always a regular point (also shown in \cite[Proposition 3.1]{Connelly05}). Then we have $\textrm{rank}(R_w(p))=\kappa$. It follows from \cite[Proposition 2]{Asimow78} that there exists a neighborhood $U_{p}$ of $p$, such that $r_\mathcal{G}^{-1}(r_{\mathcal{G}}(p))\cap U_{p}$ is a manifold of dimension $nd-\kappa$. By Corollary \ref{co gk=rk} and \cite[Theorem]{Asimow78}, $r_\mathcal{K}^{-1}(r_\mathcal{K}(p))=g_\mathcal{K}^{-1}(g_\mathcal{K}(p))$ is a manifold of dimension $(a+1)(2d-a)/2$, where $a=\textrm{rank}(p_1,\cdots,p_n)$. Together with the fact that $(\mathcal{G},p)$ is weakly rigid for $\mathcal{T}_{\mathcal{G}}^*$, there must hold $nd-\kappa=(a+1)(2d-a)/2$.

Note that there must hold $a=d$ for the generic configuration $p$, otherwise the determinant of each $d\times d$ minor of $P=(p_1,\cdots,p_n)$ is zero, which conflicts with algebraic independence of $p$. It follows that $\textrm{rank}(R_w(p))=\kappa=nd-d(d+1)/2$. That is, $(\mathcal{G},p)$ is infinitesimally weakly rigid.
\end{proof}

From Theorem \ref{th girtogr}, infinitesimal weak rigidity implies weak rigidity. Together with Theorems \ref{th ir is generic} and \ref{th r with generic to ir}, it is natural to obtain the following result.
\begin{theorem}\label{th generically weakly rigid}
If $(\mathcal{G},p)$ is weakly rigid for a generic configuration $p\in\mathbb{R}^{nd}$, then graph $\mathcal{G}$ is generically weakly rigid in $\mathbb{R}^d$.
\end{theorem}

By Theorem \ref{th ns condition for GIS}, it is straightforward that any connected graph is generically weakly rigid in the plane. Moreover, Theorem \ref{th generically weakly rigid} implies that for a generically weakly rigid graph $\mathcal{G}$ in $\mathbb{R}^d$, by randomizing a configuration $p\in\mathbb{R}^{nd}$, $(\mathcal{G},p)$ is weakly rigid with probability 1.

\section{Application to formation control}
\label{sec:app}

In this section, we aim to design distributed control laws for a multi-agent system to solve the formation stabilization problem. The desired formation shape will be characterized by using weak rigidity theory. Since we have shown that weak rigidity requires fewer edges to recognize a framework, the restriction on the formation graph will be relaxed compared to \cite{Krick09,Oh14a,Zelazo15,Sun15,Sun16auto,Sun16scl,Aranda16,Lin16}.

\subsection{Control objective}

Consider $n$ autonomous agents moving in $\mathbb{R}^d$. In a given global coordinate frame, we denote the position of agent $i$ as
$p_i\in\mathbb{R}^d$. Each agent is considered to have single-integrator dynamics:
\begin{equation}\label{dynamics}
\dot{p}_i=u_i,   ~~~~i\in\mathcal{V},
\end{equation}
where $u_i\in\mathbb{R}^d$ is a velocity input to be designed.

We denote the formation shape by $(\mathcal{G}_f,p^*)$ with $\mathcal{G}_f=(\mathcal{V},\mathcal{E}_f)$, where $\mathcal{G}_f$ is called the formation graph, and $p^*=(p_1^{*T},\cdots,p_n^{*T})^T\in\mathbb{R}^{nd}$ is a configuration forming the desired formation shape. We represent the sensing graph by $\mathcal{G}_s=(\mathcal{V},\mathcal{E}_s)$,  which describes the interaction relationships between agents. It is natural to assume that $\mathcal{E}_f\subseteq\mathcal{E}_s$ since the desired edge information is useless in design of the control input if the involved agents are unable to interact with each other. Let $\mathcal{N}_i^f$ and $\mathcal{N}_i^s$ denote the neighbor sets of agent $i$ in $\mathcal{G}_f$ and $\mathcal{G}_s$, respectively. It is easy to see that $\mathcal{N}_i^f\subseteq\mathcal{N}_i^s$.

We always consider that all agents are in a GPS-denied environment. Each agent $i$ can only achieve the relative position measurements from its neighbors in graph $\mathcal{G}_s$, i.e., $p_i^i-p_j^i$, $(i,j)\in\mathcal{E}_s$, where $p_k^i$ denotes the position vector of agent $k$ in the local coordinate system of agent $i$.

Different from the distance-constrained formation control strategies, we encode the target formation $(\mathcal{G}_f,p^*)$ through a constraint set of pairwise inner products of relative position states, i.e., $\{(p_i^*-p_j^*)^T(p_i^*-p_k^*)\in\mathbb{R}:(i,j,k)\in\mathcal{T}_{\mathcal{G}_f}^*\}$. If $(\mathcal{G}_f,p^*)$ is infinitesimally weakly rigid, then the desired equilibrium can be described by the $d(d+1)/2$ dimensional manifold $$\mathscr{E}=r_{\mathcal{K}}^{-1}r_{\mathcal{K}}(p^*)=\{p\in\mathbb{R}^{nd}: (p_i-p_j)^T(p_i-p_k)=(p_i^*-p_j^*)^T(p_i^*-p_k^*), i,j,k\in\mathcal{V}\}.$$
Remark \ref{re congruence} has shown that all agents with position states in $\mathscr{E}$ form the desired formation shape. As a result, our objective is to design distributed control strategies $u_i$ for stabilizing agents' position states into $\mathscr{E}$ asymptotically.

A framework $(\mathcal{G},p)$ is said to be realizable with $\mathcal{T}_{\mathcal{G}}^*$ if there exists some $q\in\mathbb{R}^{nd}$ such that $r_{\mathcal{G}}(q)=r_{\mathcal{G}}(p)$. Throughout this paper, we always assume that the framework characterizing the desired formation shape is realizable. The weak rigidity based formation stabilization problem is formally stated below.

\begin{problem}
Given a realizable infinitesimally weakly rigid formation $(\mathcal{G}_f,p^*)$, design a distributed control protocol (\ref{dynamics}) for each agent $i$ based on the relative position measurements $\{p^i_i-p^i_j,~j\in\mathcal{N}_i^s\}$, such that the trajectories of agents asymptotically converge into manifold $\mathscr{E}$.
\end{problem}

\subsection{A distributed control law}

Given a framework $(\mathcal{G}_f,p^*)$ and $\mathcal{T}_\mathcal{G}^*$ describing the desired formation shape, let $\delta_{(i,j,k)}=e_{ij}^Te_{ik}-e_{ij}^{*T}e_{ik}^*$, $(i,j,k)\in\mathcal{T}_\mathcal{G}^*$. We aim to steer agents to cooperatively minimize the following cost function:
\begin{equation}\label{costfun}
V(p)=\frac12\sum_{(i,j,k)\in\mathcal{T}_{\mathcal{G}_f}^*}(e_{ij}^Te_{ik}-e_{ij}^{*T}e_{ik}^*)^2 =\frac12\sum_{(i,j,k)\in\mathcal{T}_{\mathcal{G}_f}^*}\delta_{(i,j,k)}^2 ,
\end{equation}
where $e_{ij}^*=p_i^*-p_j^*$. On the basis of function (\ref{costfun}), a gradient based control law can be induced as
\begin{equation}\label{gprotocol}
u_i=-\sum_{(j,k)\in\mathcal{N}_{\mathcal{T}_i}^f}(e_{ij}^Te_{ik}-e_{ij}^{*T}e_{ik}^*)(e_{ij}+e_{ik}) -\sum_{(j,k)\in\mathcal{N}_{\mathcal{T}^i}^f}(e_{ji}^Te_{jk}-e_{ji}^{*T}e_{jk}^*)(e_{ij}-e_{ik}),
\end{equation}
where $\mathcal{N}_{\mathcal{T}_i}^f=\{(j,k)\in\mathcal{V}\times\mathcal{V}:(i,j,k)\in\mathcal{T}_{\mathcal{G}_f}^*\}$, $\mathcal{N}_{\mathcal{T}^i}^f=\{(j,k)\in\mathcal{V}\times\mathcal{V}:(j,i,k)\in\mathcal{T}_{\mathcal{G}_f}^*~or~ (j,k,i)\in\mathcal{T}_{\mathcal{G}_f}^*\}$.

Note that for a triple $(i,j,k)\in\mathcal{T}_{\mathcal{G}_f}^*$, the gradient of $\delta_{(i,j,k)}$ with respect to $p_j$ always involves the information of $e_{ik}$. That is, agent $j$ should have access to $e_{ik}$, which can be computed by $e_{ij}-e_{kj}$. This implies that agent $j$ should be able to sense information from $i$ and $k$. Similarly, agent $k$ should be able to sense information from $i$ and $j$. Therefore, we require that $(i,j),(j,k),(i,k)\in\mathcal{E}_s$. To implement the control law (\ref{gprotocol}) distributively, the set $\mathcal{T}_{\mathcal{G}_f}^*$, which includes the target inner products of displacements, is constructed by
\begin{equation}\label{T_{G_f}}
\mathcal{T}_{\mathcal{G}_f}^*=\{(i,j,k)\in\mathcal{V}^3: (i,j),(i,k)\in\mathcal{E}_f, ~j=k~or~ (j,k)\in\mathcal{E}_s, j\leq k\}.
\end{equation}

\begin{remark}
Observe that before implementing the control law (\ref{gprotocol}), each agent $i$ should be assigned with the elements of $\mathcal{T}_{\mathcal{G}_f}^*$ involving $i$ and the target inner product constraints involving $i$. This can be viewed as a centralized distribution \cite{Summers11}, which is similar to displacement- or distance-based strategies. When the decentralized controller (\ref{gprotocol}) is implemented, each agent $i$ only has to sense relative displacements from neighbors and compute the control input by simple inner products, additions and subtractions of vectors. Therefore, the control law (\ref{gprotocol}) is reasonable from the practical point of view.
\end{remark}

We make the following assumption on the target formation $(\mathcal{G}_f,p)$.
\begin{assumption}\label{assum1}
$(\mathcal{G}_f,p)$ with $\mathcal{T}_{\mathcal{G}_f}^*$ is infinitesimally weakly rigid.
\end{assumption}

In this paper, infinitesimal weak rigidity of $(\mathcal{G}_f,p)$ is the only condition for solving the formation stabilization problem. In fact, since $\mathcal{T}_{\mathcal{G}_f}^*$ is constructed as (\ref{T_{G_f}}), to satisfy Assumption \ref{assum1}, $\mathcal{G}_s$ may be required to have more edges than $\mathcal{G}_f$. More specifically, for $(i,j,k)\in\mathcal{T}_{\mathcal{G}_f}^*$ and $j\neq k$, it should hold that $(j,k)\in\mathcal{E}_s$, but it is unnecessary that $(j,k)\in\mathcal{E}_f$. As an example, for the sensing graph shown in Fig. \ref{fig 4 examples} (a), the edge $(2,3)$ can be reduced to generate the formation graph in Fig. \ref{fig 4 examples} (b). It is important to note that using (\ref{T_{G_f}}) as $\mathcal{T}_{\mathcal{G}_f}^*$ is mainly to guarantee effectiveness of the gradient controller (\ref{gprotocol}). As a result, the restriction of $\mathcal{G}_s$ can be relaxed if some other distributed controller is applied. We will introduce the detail in Subsection \ref{subsection3}.

The following lemma shows the implicit condition on $\mathcal{G}_s$ for validity of Assumption \ref{assum1}.
\begin{lemma}\label{le condition for G_s}
Assumption \ref{assum1} holds if and only if $(\mathcal{G}_s,p)$ is infinitesimally rigid.
\end{lemma}
\begin{proof} Let $r_{\mathcal{G}_f}=(\cdots,e_{ij}^Te_{ik},\cdots)^T,(i,j,k)\in\mathcal{T}_{\mathcal{G}_f}^*$, $R_w=\frac{\partial r_{\mathcal{G}_f}}{\partial p}=(\xi_1,\cdots,\xi_s)^T$, $g_{\mathcal{G}_s}=(\cdots,||e_{ij}||^2,\cdots)^T,(i,j)\in\mathcal{E}$ be the distance-based rigidity function of $(\mathcal{G}_s,p)$, $R=\frac{\partial g_{\mathcal{G}_s}}{\partial p}$. It follows from (\ref{T_{G_f}}) that $\mathcal{T}_{\mathcal{G}_f}^*$ always includes $\{(i,j,j):(i,j)\in\mathcal{E}_s\}$, implying that $\textrm{rank}(R_w)\geq \textrm{rank}(R)=nd-\frac{d(d+1)}{2}$. Therefore the sufficiency is obtained. Next we prove the necessity.

It suffices to show that each row of $R_w$ can be denoted by a linear combination of rows of $R$. It is obvious that we only have to focus on $\xi_l^T=\frac{\partial (e_{ij}^Te_{ik})}{\partial p}$ for $j\neq k$. Note that it always holds $e_{ij}^Te_{ik}=(||e_{ij}||^2+||e_{ik}||^2-||e_{jk}||^2)/2$, it follows that $\frac{\partial (e_{ij}^Te_{ik})}{\partial p}=(\frac{\partial (||e_{ij}||^2)}{\partial p}+\frac{\partial (||e_{ik}||^2)}{\partial p}-\frac{\partial (||e_{jk}||^2)}{\partial p})/2$, implying that each row in $R_w$ can be denoted by several rows of $R$. Therefore $\textrm{rank}(R)\geq \textrm{rank}(R_w)=nd-d(d+1)/2$. Recall that $\textrm{null}(R)\geq d(d+1)/2$, we have $\textrm{rank}(R)=nd-d(d+1)/2$.
\end{proof}

It is worthwhile to note that in a particular case when $\mathcal{G}_s$ is a triangulated Laman graph (See details in \cite{Chen17}), $(\mathcal{G}_s,p)$ can be a minimally infinitesimally rigid framework. Then Assumption \ref{assum1} holds from Lemma \ref{le condition for G_s}.

Let $\delta(p)=(\cdots,\delta_{(i,j,k)},\cdots)^T=r_{\mathcal{G}_f}(p)-r_{\mathcal{G}_f}(p^*)$, then $V(p)=\frac12||\delta(p)||^2$. By the chain rule, the dynamic equation of multi-agent system (\ref{dynamics}) with control law (\ref{gprotocol}) can be written in the following compact form
\begin{equation}\label{cprotocol}
\dot{p}=-(\nabla_{p}V(p))^T=-(\delta^T(p)\cdot\frac{\partial r_{\mathcal{G}_f}(p)}{\partial p})^T=-R_w^T(p)\delta(p).
\end{equation}
Under Assumption \ref{assum1}, each agent can achieve the required information for implementing controller (\ref{gprotocol}) via local interactions with its neighbors. Therefore our control strategy is a distributed one. In fact, we also have the following properties for the control law (\ref{gprotocol}).
\begin{lemma}\label{le property}

(i) The controller (\ref{gprotocol}) is independent of the global coordinate frame.

(ii) The centroid $\bar{p}=\frac{1}{n}\sum_{i\in\mathcal{V}}p_i(t)$ is invariant, i.e., $\dot{\bar{p}}=0$.

(iii) Let $P=(p_1,\cdots,p_n)$, $\textrm{rank}(P(0))=\textrm{rank}(P(t))$ for all $t\geq0$.
\end{lemma}
\begin{proof} (i) and (ii) are straightforward by a proof similar to \cite{Oh11}, thus we omit the proofs here.

(iii) Let $R_e=\frac{\partial r_{\mathcal{G}_f}}{\partial e}\in\mathbb{R}^{s\times {md}}$ with $e=(\cdots,e_{ij}^T,\cdots)^T=(\alpha_1^T,\cdots,\alpha_m^T)^T\in\mathbb{R}^{md}$, and $e^*=(\cdots,e_{ij}^{*T},\cdots)^T=(\alpha_1^{*T},\cdots,\alpha_m^{*T})^T\in\mathbb{R}^{md}$. Note that we always have
$$R_e^T\delta=(\triangle\otimes I_d)e,$$
where $\delta=(\delta_1,\cdots,\delta_s)^T=e^Te-e^{*T}e^*$, $\triangle=[\triangle_{ij}]\in\mathbb{R}^{m\times m}$,  $\triangle_{ii}=||\alpha_i||^2-||\alpha_i^*||^2$ if $||\alpha_i||^2-||\alpha_i^*||^2$ is a component of $\delta$, and $\triangle_{ii}=0$ otherwise; $\triangle_{ij}=\triangle_{ji}=\alpha_i^T\alpha_j-\alpha_i^{*T}\alpha_j^*$ if $\alpha_i^T\alpha_j-\alpha_i^{*T}\alpha_j^*$ is a component of $\delta$, and $\triangle_{ij}=\triangle_{ji}=0$ otherwise. As a result, (\ref{cprotocol}) is transformed to $$\dot{p}=-R_w^T\delta=-\bar{H}^TR_e^T\delta=-\bar{H}^T(\triangle\otimes I_d)\bar{H}p= -((H^T\triangle H)\otimes I_d)p,$$
which can be equivalently written as
$$\dot{P}=-P\bar{\triangle},$$
where $\bar{\triangle}=H^T\triangle H$. From the lemma on rank-preserving differential equations shown in \cite{Sun15}, we can obtain that $\textrm{rank}(P)$ is invariant during evolution of the formation system.
\end{proof}

\subsection{Stability analysis}

Notice that the cost function (\ref{costfun}) is nonconvex, thus has multiple local minima. We will mainly concentrate on local stability of the formation system (\ref{cprotocol}). We say a set $M$ is locally exponentially stable if there exists an exponent $c>0$, such that for any $x\in M$, there exists a neighborhood $\Omega$ of $x$, any trajectory starting from $\Omega$ converges to $M$ at least as fast as $e^{-ct}$.

\begin{theorem}\label{th exponential stable}
For a group of $n> d+1$ agents with dynamics (\ref{dynamics}) and control law (\ref{gprotocol}) moving in $\mathbb{R}^d$, under Assumption \ref{assum1}, $\mathscr{E}$ is locally exponentially stable.
\end{theorem}
\begin{proof} We observe that the gradient based formation system (\ref{cprotocol}) has a similar form to \cite[Equation (8)]{Krick09}. Let $z=(\bar{p}^T,\bar{z}^T)^T=Qp$, where $\bar{z}\in\mathbb{R}^{nd-d}$, $Q\in\mathbb{R}^{nd\times nd}$ is an orthogonal matrix with its first $d$ rows being $\frac{1}{n}\mathbf{1}_n^T\otimes I_d$. Using the centroid invariance property, i.e., $\dot{\bar{p}}=0$, shown in Lemma \ref{le property}, (\ref{cprotocol}) can also be equivalently transformed into a reduced-order system $\dot{\bar{z}}=\bar{f}(\bar{z})$ with a compact manifold of equilibria $\bar{\mathscr{E}}$. Since the target formation $\mathscr{E}=r^{-1}_{\mathcal{K}}(r_{\mathcal{K}}(p^*))$ has been shown to be a $d(d+1)/2-$dimensional manifold in the proof of Theorem \ref{th girtogr}, $\bar{\mathscr{E}}$ is a $d(d-1)/2-$dimensional manifold characterized by rotations around the centroid $\bar{p}(0)$. 
	
By simply following a similar procedure to that in \cite{Krick09}, the traditional center manifold theory can be employed to show that each point in $\bar{\mathscr{E}}$ is locally exponentially stable. Due to compactness of $\bar{\mathscr{E}}$, there is a finite subcover forming a neighborhood of $\bar{\mathscr{E}}$. Therefore, there must exist an exponent $c>0$ such that for each $\bar{z}\in\bar{\mathscr{E}}$, any trajectory converges to $\bar{\mathscr{E}}$ from a neighborhood of $\bar{z}$ at least as fast as $e^{-ct}$. Recall that $\bar{p}$ is invariant, it is straightforward that the same conclusion holds for $\mathscr{E}$.
\end{proof}

Note that although the desired equilibrium is $\mathscr{E}$, and $r_{\mathcal{G}_f}^{-1}(r_{\mathcal{G}_f}(p^*))=\mathscr{E}$ only if $(\mathcal{G}_f,p)$ is globally weakly rigid, the target formation is not necessary to be globally weakly rigid since we only require local stability.

Local exponential stability actually characterizes the ability of control law (\ref{gprotocol}) to restore the desired formation shape under a small perturbation from the desired equilibrium $\mathscr{E}$. In fact, when $n=d+1$ and $(\mathcal{G}_f,p^*)$ with $\mathcal{T}_{\mathcal{G}_f}^*$ is minimally infinitesimally weakly rigid, almost global asymptotic stability of the formation system can be ensured, as given in the following theorem.

\begin{theorem}\label{th n=d+1}
For a group of $n=d+1$ agents with dynamics (\ref{dynamics}) and control law (\ref{gprotocol}) moving in $\mathbb{R}^d$, if $p_1(0),\cdots,p_n(0)$ do not lie in a hyperplane and $(\mathcal{G}_f,p^*)$ with $\mathcal{T}_{\mathcal{G}_f}^*$ is minimally infinitesimally weakly rigid, then

(i) $(\mathcal{G}_f,p(t))$ is infinitesimally weakly rigid for all $t\geq0$;

(ii) Collisions between any agents are avoided;

(iii)  The stacked state vector $p$ will converge into $\mathscr{E}$ exponentially.
\end{theorem}
\begin{proof} (i)
According to Lemma \ref{le condition for G_s}, $\textrm{rank}(R(p(0)))=\textrm{rank}(R_w(p(0)))=nd-d(d+1)/2$, where $R(p(t))=\frac{\partial g_{\mathcal{G}_s}}{\partial p}$. It follows that $|\mathcal{E}_s|\geq nd-d(d+1)/2=n(n-1)/2=C_n^2$. Thus $\mathcal{G}_s$ is a complete graph. By Lemma \ref{le condition for G_s}, we only have to prove that $R(p(t))$ is of full row rank for $t\geq0$. Suppose this is not true for $t=t^*$. Let $R(p(t^*))=(r_1,\cdots,r_{\bar{m}})^T=(c_1,\cdots,c_n)$, where $r_i\in\mathbb{R}^{nd}$, $i\in\{1,\cdots,\bar{m}\}$, $c_j\in\mathbb{R}^{\bar{m}\times d}$, $j\in\mathcal{V}$, $\bar{m}=n(n-1)/2=d(d+1)/2$. Then there exist not all zero scalars $\tau_1,\cdots,\tau_{\bar{m}}$ such that $\tau_1r_1+\cdots+\tau_{\bar{m}}r_{\bar{m}}=0$, which implies that $\tau^TR(p(t^*))=\tau^T(c_1,\cdots,c_n)=0$. Without loss of generality, suppose $\tau_k\neq0$. Note that there must exist a submatrix $c_l$ such that the $k$th row of $c_l$ is associated with $e_{li}^T(t^*)$ for some $i\in\mathcal{N}_l^s$. Since $\mathcal{G}_s$ is complete, $c_l=(\frac{\partial g_{\mathcal{G}_s}}{\partial p_l})^T$ has exactly $n-1=d$ nonzero rows associated with $e_{li}^T(t^*)$ for $i\in\mathcal{N}_l^s$. Together with the fact that $\tau^Tc_l=0$, where $\tau=(\tau_1,\cdots,\tau_{\bar{m}})^T$. These $e_{li}(t^*)$ for all $i\in\mathcal{N}_l^s$ are linearly dependent. Let $E_l=(\cdots,e_{li}(t^*),\cdots)^T\in\mathbb{R}^{d\times d},i\in\mathcal{N}_l^s$, we have $\textrm{rank}(E_l)<d$.

Since $p_1(0),\cdots,p_n(0)$ do not lie on a hyperplane, there does not exist not all zero scalars $k_1,\cdots,k_d$ and $b\in\mathbb{R}$ such that $k_1p_{i(1)}(0)+\cdots+k_dp_{i(d)}(0)=b$ for all $i\in\mathcal{V}$, where $p_{i(j)}(0)\in\mathbb{R}$ denotes the $j$th component of $p_i(0)$. As a result, the matrix $\bar{P}(0)=(\mathbf{1}_n,P^T(0))\in\mathbb{R}^{(d+1)\times(d+1)}$ is of full rank. Due to Lemma \ref{le property}, one has $\dot{P}^T=-\bar{\triangle}^TP^T=-\bar{\triangle}P^T$, together with the fact that $\bar{\triangle}\mathbf{1}_n=H^T\triangle H\mathbf{1}_n=0$, we have $\dot{\bar{P}}=-\bar{\triangle}\bar{P}$, which is a rank preserving differential equation. Hence, $\bar{P}(t^*)$ is of full rank, and $P^T(t^*)$ is of full column rank. Since $\mathcal{G}_s$ is complete, we have $E_l=H_lP^T(t^*)$, where $H_l$ is an incidence matrix associated with a star topology with agent $l$ as the root. Note that $\textrm{null}(H_l)=\textrm{span}\{\mathbf{1}_n\}$, if $E_lx=0$ for some nontrivial $x\in\mathbb{R}^{nd}$, then either $P^Tx=0$ or $p_i^Tx=p_j^Tx$ for all $i,j\in\mathcal{V}$. Both the two cases cannot happen since $P^T(t^*)$ is of full column rank and $\bar{P}(t^*)$ is of full rank. As a result, $\textrm{null}(E_l)=\varnothing$. This conflicts with $\textrm{rank}(E_l)<d$. Consequently, we have $\textrm{rank}(R_w(p(t)))=\textrm{rank}(R(p(t)))=nd-d(d+1)/2$ for all $t\geq0$.

(ii) Suppose that there are two agents $i,j$ colliding with each other at some $t\geq0$. From (i) and Lemma \ref{le collision avoidance}, agent $i$ should have $d$ neighbors other than $j$. It follows that $|\mathcal{N}^s_i|\geq d+1>d$, which conflicts with $|\mathcal{N}^s_i|=d$. Hence, collision avoidance is guaranteed during the formation process.

(iii) We first claim that $r_{\mathcal{G}_f}$ has exactly $\bar{m}=nd-d(d+1)/2=d(d+1)/2$ components. It follows from Theorem \ref{th ns condition for GIS} that $\mathcal{G}_f$ is a tree. From the form of $\mathcal{T}_{\mathcal{G}_f}^*$ given by (\ref{T_{G_f}}), we have $|\mathcal{T}_{\mathcal{G}_f}^*|=n-1+C_{n-1}^2=d(d+1)/2=\bar{m}$, therefore, $r_{\mathcal{G}_f}\in\mathbb{R}^{\bar{m}}$. Next we show $r_{\mathcal{G}_f}(p)=r_{\mathcal{G}_f}(p^*)$ implies $p\in\mathscr{E}$, i.e., $(\mathcal{G}_f,p)$ is globally weakly rigid. Note that it always holds $e_{ij}^Te_{ik}=1/2(||e_{ij}||^2+||e_{ik}||^2-||e_{jk}||^2)$, i.e., any component of $r_{\mathcal{G}_f}$ can be denoted by a linear combination of several components of $g_{\mathcal{G}_s}$. Therefore, there exists a constant matrix $M\in\mathbb{R}^{\bar{m}\times\bar{m}}$ such that $r_{\mathcal{G}_f}(p)=Mg_{\mathcal{G}_s}(p)$ for any $p\in\mathbb{R}^{nd}$. It follows that $R_w=\frac{\partial r_{\mathcal{G}_f}}{\partial p}=M\frac{\partial g_{\mathcal{G}_s}}{\partial p}=MR$. From the well known inequality $\textrm{rank}(M)+\textrm{rank}(R)-\bar{m}\leq \textrm{rank}(R_w)\leq \min\{\textrm{rank}(M),\textrm{rank}(R)\}$, we have $\textrm{rank}(M)=\bar{m}$, i.e., $M$ is nonsingular. It follows that $g_{\mathcal{G}_s}=M^{-1}r_{\mathcal{G}_f}$. That is, once $r_{\mathcal{G}_f}(p)=r_{\mathcal{G}_f}(p^*)$, it holds that $g_{\mathcal{G}_s}(p)=g_{\mathcal{G}_s}(p^*)$. Recall that $\mathcal{G}_s$ is complete, we have $D(p)=D(p^*)$. According to Theorem \ref{th D=E}, it holds that $\mathbf{E}(p)=\mathbf{E}(p^*)$, i.e., $p\in\mathscr{E}$.

Next we prove exponential stability of $\{p\in\mathbb{R}^{nd}:r_{\mathcal{G}_f}(p)=r_{\mathcal{G}_f}(p^*)\}$. Let $\delta=r_{\mathcal{G}_f}(p)-r_{\mathcal{G}_f}(p^*)$. Due to the fact that $\dot{\delta}=\dot{r}_{\mathcal{G}_f}=\frac{\partial r_{\mathcal{G}_f}}{\partial p}\dot{p}=R_w\dot{p}$, together with (\ref{cprotocol}), the formation system can be described by
\begin{equation}\label{dequation}
\dot{\delta}=-R_wR_w^T\delta.
\end{equation}
Now we show (\ref{dequation}) is a self-contained system. It suffices to show $R_wR_w^T$ is a function of $\delta$. Note that each entry of $R_wR_w^T$ is a linear combination of $e_{ij}^Te_{kl}$, $(i,j),(k,l)\in\mathcal{E}$. It is easy to see that $e_{ij}^Te_{kl}$ can always be denoted by a linear combination of components of $g_{\mathcal{G}_s}$, i.e., $e_{ij}^Te_{kl}=\frac12(||e_{jk}||^2+||e_{il}||^2-||e_{ik}||^2-||e_{jl}||^2)$. Together with $g_{\mathcal{G}_s}=M^{-1}r_{\mathcal{G}_f}$, it becomes certain that $R_wR_w^T$ can be written as a smooth function of $r_{\mathcal{G}_f}$, therefore is also a smooth function of $\delta$.

Let $\phi=||\delta||^2$, it follows from (\ref{dequation}) that $\dot{\phi}=-2\delta^TR_wR_w^T\delta\leq0$. This implies that $\delta$ always stays in the compact set $\Psi=\{\delta\in\mathbb{R}^{\bar{m}}:||\delta||^2\leq\phi(0)\}$. Since we have shown in (i) that $\textrm{rank}(R_wR_w^T)=\textrm{rank}(R_w)=\bar{m}$ for all $t\geq0$, together with the fact that $R_wR_w^T$ is totally determined by $\delta$, there must exist $\kappa>0$ such that $\kappa =\min_{\delta\in\Psi}\lambda(R_wR_w^T)$. It follows that $\dot{\phi}\leq -2\kappa\phi$. Then $\phi\leq \exp(-2\kappa)\phi(0)$. That is, $\delta$ vanishes exponentially.

Recall that $(\mathcal{G}_f,p)$ is globally weakly rigid, hence, $p$ must converge into $\mathscr{E}$ exponentially.
\end{proof}

\begin{remark}
In the case when $n\geq d+1$, let $\mathcal{C}$ be the set of configurations in a hyperplane of $\mathbb{R}^d$. Then $\mathcal{C}=\{p\in\mathbb{R}^{nd}:f(p)=0\}$, where $f(p)$ is the sum of the squares of all the $(d+1)\times(d+1)$ minors of $\bar{P}=(\mathbf{1}_n,P^T)\in\mathbb{R}^{n\times(d+1)}$. Note that $f(p)$ is a nontrvial polynomial, thus $\mathcal{C}$ is either equal to $\mathbb{R}^{nd}$ or of measure zero \cite{Caron05}. Since $p^*$ is a configuration such that $\textrm{rank}(\bar{P})=d+1$, we have $p^*\notin\mathcal{C}$. Therefore, the measure of $\mathcal{C}$ is zero. That is, if $n=d+1$, Theorem \ref{th n=d+1} implies that for almost any given initial configuration, the control law (\ref{gprotocol}) will exponentially stabilize a minimally infinitesimally weakly rigid formation. However, since the exponent $2\kappa$ is dependent on $p(0)\in\mathbb{R}^{nd}\setminus\mathcal{C}$, and $\mathbb{R}^{nd}\setminus\mathcal{C}$ is not compact, it is uncertain that whether a uniform $\kappa$ exists. Thus, we can only conclude that if $n=d+1$, the minimally infinitesimally weakly rigid formation is almost globally asymptotically stable.
\end{remark}


\subsection{Formation control under non-rigid sensing graphs}\label{subsection3}

Lemma \ref{le condition for G_s} shows that the control strategy proposed in previous subsections can only be implemented on the premise that $(\mathcal{G}_s,p^*)$ is infinitesimally rigid. In this subsection, we consider $\mathcal{G}_f=\mathcal{G}_s=\mathcal{G}=(\mathcal{V},\mathcal{E})$ and try to solve the formation stabilization problem when $(\mathcal{G},p^*)$ is only infinitesimally weakly rigid, which is a weaker graph condition compared to rigidity. Different from (\ref{T_{G_f}}), we use $\mathcal{T}_{\mathcal{G}}^*$ as follows.

\begin{equation}\label{T_G}
\mathcal{T}_{\mathcal{G}}^*=\{(i,j,k)\in\mathcal{V}^3: (i,j),(i,k)\in\mathcal{E}, j\leq k\}.
\end{equation}
To achieve the goal of formation, we consider the following independent cost function for each agent:
\begin{equation}\label{nonrigid costfun}
V_i(p)=\frac12\sum_{(j,k)\in\mathcal{N}_{\mathcal{T}_i}}\delta_{(i,j,k)}^2+ \frac12\sum_{j\in\mathcal{N}_i}\delta_{(j,i,i)}^2, ~~~~~~i\in\mathcal{V},
\end{equation}
where $\mathcal{N}_{\mathcal{T}_i}=\{(j,k)\in\mathcal{V}\times\mathcal{V}:(i,j,k)\in\mathcal{T}_{\mathcal{G}}^*\}$.

The distributed control law is
\begin{equation}\label{nonrigid protocol}
u_i=-K_i\nabla_{p_i}V_i, ~~~~~~i\in\mathcal{V},
\end{equation}
where $K_i\in\mathbb{R}^{d\times d}$ is a control gain matrix for agent $i$ to be designed. Note that the control law (\ref{nonrigid protocol}) for agent $i$ induced by function (\ref{nonrigid costfun}) only requires information of $p_i-p_j$ for $(i,j)\in\mathcal{E}_f$. Hence a sensing graph $\mathcal{G}_s=\mathcal{G}_f$ is sufficient for each agent to implement the control law. Moreover, (\ref{nonrigid protocol}) is no longer a normal gradient based control law. As a result, the centroid $\bar{p}$ is dynamic during the formation process. However, it can be easily verified that the control law (\ref{nonrigid protocol}) is still independent of the global coordinate frame.

With (\ref{nonrigid protocol}), the formation system can be written as
\begin{equation}\label{nonrigid formation system}
\dot{p}=-K\bar{R}_w^T\delta,
\end{equation}
where $K=\textrm{diag}(K_1,\cdots,K_n)\in\mathbb{R}^{nd\times nd}$.

Different from (\ref{cprotocol}), the non-gradient based formation system (\ref{nonrigid formation system}) has a dynamic centroid, thus cannot be transformed to a reduced-order system with a compact manifold of equilibria. Due to non-compactness of $\mathcal{E}$, there does not exist an open cover of $\mathcal{E}$ having a finite subcover. In the following, we will employ Lemma \ref{le manifold theory} to establish local exponential stability of the target formation shape.

\begin{theorem}\label{th nonrigid formation}
For a group of $n\geq d+1$ agents with dynamics (\ref{dynamics}) and control law (\ref{nonrigid protocol}) moving in $\mathbb{R}^d$, under Assumption \ref{assum1}, if for some $p^*\in\mathscr{E}$, the gain matrix $K$ can be chosen such that $J^*=K\bar{R}_w^TR_w|_{p=p^*}$ has $d(d+1)/2$ zero eigenvalues and the rest have positive real parts, then for any $\tilde{p}\in\mathscr{E}$, there exists a compact neighborhood $\Omega$ of $\tilde{p}$, such that $M=\Omega\cap\mathscr{E}$ is locally exponentially stable.
\end{theorem}
\begin{proof} For any $\tilde{p}\in\mathscr{E}$, let $\rho=p-\tilde{p}$. Rewrite (\ref{nonrigid formation system}) as $\dot{p}=f(p)$, expanding in a Taylor series about $\tilde{p}$, we have $f(p)=f(\tilde{p})+\frac{\partial f(\tilde{p})}{\partial p}\rho+g(\rho)$. Due to the fact that $\mathscr{E}$ is a manifold of equilibria, we have $\dot{\tilde{p}}=f(\tilde{p})$. Then (\ref{nonrigid formation system}) can be equivalently written as
\begin{equation}\label{rho dynamics}
\dot{\rho}=\frac{\partial f(\tilde{p})}{\partial p}\rho+g(\rho)=-J_f(\tilde{p})\rho+g(\rho),
\end{equation}
where $J_f(\tilde{p})=K\bar{R}_w^TR_w\big|_{p=\tilde{p}}$. Note that each entry of $\bar{R}_w^TR_w$ is either zero or an inner product of relative positions, implying that $\bar{R}_w^TR_w$ is invariant under translations and rotations of $p$. Due to the assumption, $J_f(\tilde{p})=J^*$ has $d(d+1)/2$ zero eigenvalues and the rest have positive real parts. Let $M_1=\{\rho\in\mathbb{R}^{nd}:\rho+\tilde{p}\in\mathscr{E}\}$, $M_1$ is obviously a manifold of dimension $d(d+1)/2$. Applying Lemma \ref{le manifold theory} to system (\ref{rho dynamics}), there exists a compact neighborhood $\Omega_1$ of the origin, such that $M_2=\Omega_1\cap M_1$ is locally exponentially stable. Let $\Omega=\{\tilde{p}+\rho:\rho\in \Omega_1\}$, it is straightforward that $M=\Omega\cap\mathscr{E}$ is locally exponentially stable.
\end{proof}

Observe that if $(j,k)\in\mathcal{E}$ for all $(i,j,k)\in\mathcal{T}_\mathcal{G}^*$, then $\bar{R}_w=R_w$. Let $\bar{R}_w^*=\bar{R}_w(p^*)$, $R_w^*=R_w(p^*)$. It follows that $\textrm{rank}(\bar{R}_w^{*T}R_w^*)=\textrm{rank}(R_w^*)=nd-d(d+1)/2$, and all nonzero eigenvalues of $\bar{R}_w^{*T}R_w^*$ are positive. Hence $J^*$ satisfies our condition by setting $K=I_{nd}$. Otherwise, we have $\textrm{null}(R_w^*)\subseteq \textrm{null}(\bar{R}_w^{*T}R_w^*)$, implying that $\textrm{rank}(\bar{R}_w^{*T}R_w^*)\leq nd-d(d+1)/2$. Since it always holds that $\textrm{rank}(J^*)\leq \min\{\textrm{rank}(K),\textrm{rank}(\bar{R}_w^{*T}R_w^*)\}$, to make $\textrm{rank}(J^*)=nd-d(d+1)/2$, there should hold $\textrm{rank}(\bar{R}_w^{*T}R_w^*)\geq nd-d(d+1)/2$. Therefore, a necessity condition for validity of the condition in Theorem \ref{th nonrigid formation} is $\textrm{rank}(\bar{R}_w^{*T}R_w^*)=nd-d(d+1)/2$. In particular, consider a formation stabilization problem in the plane, if $\mathcal{T}_\mathcal{G}^*$ is selected by Algorithms \ref{alg:spanning tree} and \ref{alg: construction of T_G^*}, then $\bar{R}_w^*,R_w^*\in\mathbb{R}^{(2n-3)\times2n}$. From the proof of Theorem \ref{th ns condition for GIS}, there is always one edge $(i,j)\in\mathcal{E}$ such that only two components of $r_{\mathcal{G}}$ involve $e_{ij}^*$, i.e., $||e_{ij}^*||^2$ and $e_{ij}^{*T}e_{ik}^*$ for some $k\in\mathcal{N}_i$. The corresponding two rows in $\bar{R}_w^*$ form the following submatrix

$$\bordermatrix{
             &     ~        &    i         &       ~      &     j       &      ~       &       k      &       ~     \cr
& \mathbf{0}   & 2e_{ij}^{*T}     & \mathbf{0}   &  -2e_{ij}^{*T}  & \mathbf{0}  &  \mathbf{0}   & \mathbf{0}  \cr
& \mathbf{0}   & e_{ij}^{*T}+e_{ik}^{*T}     & \mathbf{0}   &  \mathbf{0}   & \mathbf{0}  &  \mathbf{0}  & \mathbf{0}  \cr
},$$
which are obviously linearly independent. Similar to the induction in the proof of Theorem \ref{th ns condition for GIS}, it can be verified that $\textrm{rank}(\bar{R}_w^*)=\textrm{rank}(R_w^*)=2n-3$. It follows that $\textrm{rank}(\bar{R}_w^{*T}R_w^*)=2n-3$. Although numerical experiments show that a suitable $K$ can always be chosen, it is still difficult to determine the existence of $K$ rigorously and is the topic of ongoing research endeavors.

\section{A simulation example}
\label{sec:simu}

In this section, we present a numerical example to illustrate the effectiveness of the main results.

\begin{figure}
\centering
\includegraphics[width=12cm]{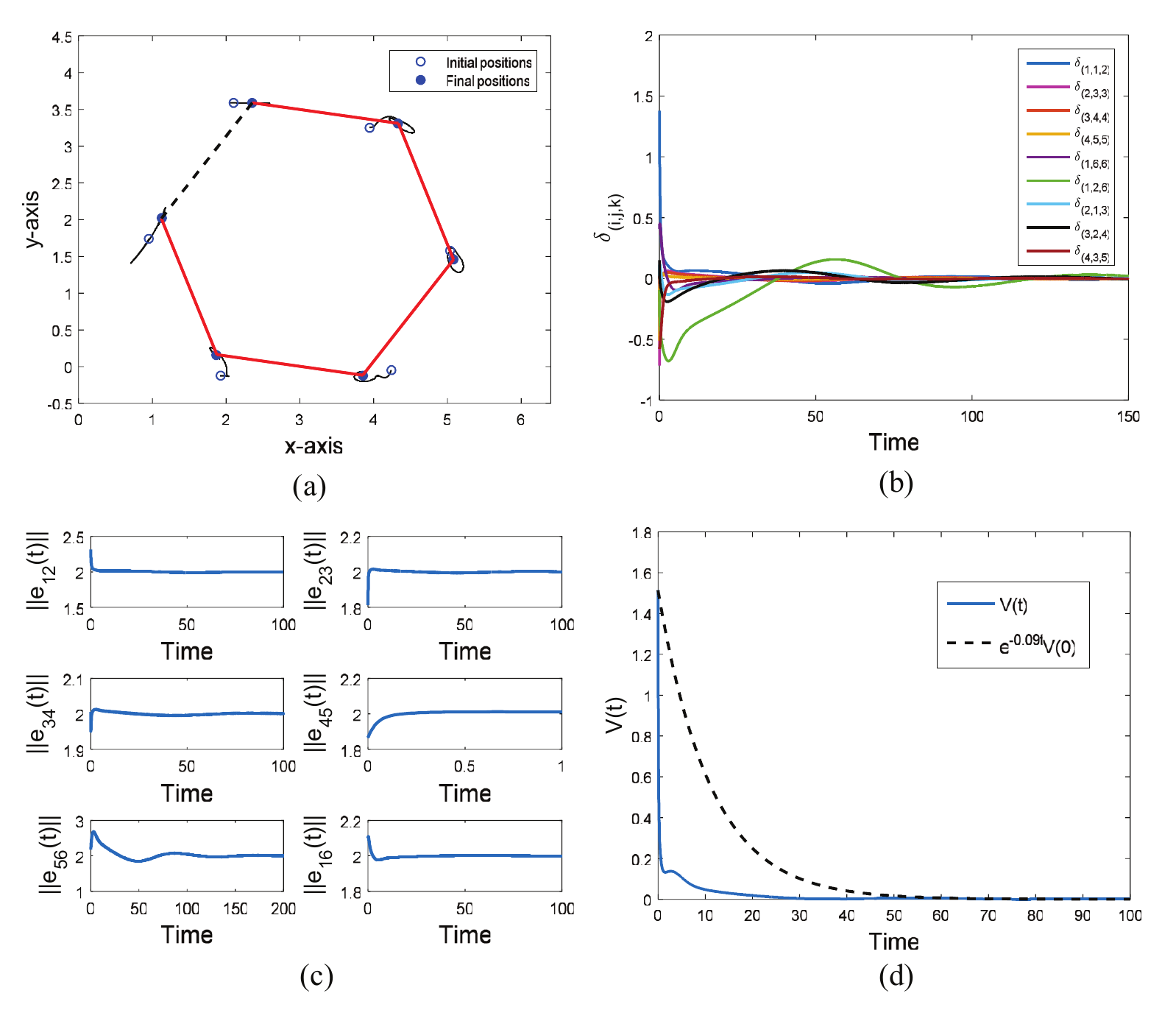}
\caption{(a) The agents with initial positions in a neighborhood of $p^*$ asymptotically converge into another point in $\mathscr{E}$. (b) $\delta_{(i,j,k)}$ asymptotically vanishes to zero, $(i,j,k)\in\mathcal{T}_\mathcal{G}^*$. (c) The length of each edge asymptotically converges to $2$. (d) The cost function $V=\sum_{i\in\mathcal{V}}V_i$ with $V_i$ in (\ref{nonrigid costfun}) vanishes to zero exponentially.}
\label{fig hexagon}
\end{figure}

Consider a group of 6 autonomous agents moving in the plane. We try to stabilize these agents to form a regular hexagon with edge lengths equal to $2$. The formation graph $\mathcal{G}_f$ and the sensing graph $\mathcal{G}_s$ are identically set to be the path graph $\mathcal{G}$ in Fig. \ref{fig conditionforR_e} (b). $(\mathcal{G},p)$ is obviously not rigid, but infinitesimally weakly rigid with $\mathcal{T}_\mathcal{G}^*=\{(1,2,6),(2,1,3),(3,2,4),(4,3,5),(i,j,j),(i,j)\in\mathcal{E},i>j\}$. Because the sensing graph is not rigid, the distance-based formation strategies are inapplicable. Now given a configuration $p^*=(2,0,4,0,5,\sqrt{3},4,2\sqrt{3},2,$ $2\sqrt{3},1,\sqrt{3})^T$ which forms the target formation shape, and let each agent implement the control law (\ref{nonrigid protocol}). In fact, if we set $K=I$, the eigenvalues of $J^*$ are 45.9712, 40.4991, 32.7903, 24.0000, 15.8549,  10.0916, 5.6563, 1.4093, -0.2727, 0, 0, 0. This does not satisfy our condition in Theorem \ref{th nonrigid formation}.   Now we employ a gain matrix $K=\textrm{diag}(K_1,\cdots,K_6)$ with $K_1=\textrm{diag}(0.3,$ $-0.04)$, $K_2=\textrm{diag}(0.15,1.34)$, $K_3=\textrm{diag}(0.23,1.09)$, $K_4=\textrm{diag}(1.32,$ $0.34)$, $K_5=\textrm{diag}(1.32,0.21)$, $K_6=\textrm{diag}(-0.45,0.42)$. Then the eigenvalues of $J^*$ become 48.9899, 36.7915, 12.6938, 8.1539, 3.7883, 2.7087, 1.7132, 0.1053+0.1757i, 0.1053-0.1757i, 0, 0, 0. By Theorem \ref{th nonrigid formation}, the desired formation shape can be formed locally exponentially, which is consistent with the result shown in Fig. \ref{fig hexagon}.

\section{Conclusion}
\label{sec:con}

We presented a weak rigidity theory, which allows us to recognize a framework in arbitrarily dimensional spaces by fewer edges than the distance-based rigidity theory. The main idea is to determine the framework by constraining pairwise inner products of relative displacements in the framework, which actually utilizes additional subtended angle information not used in  distance-based rigidity theory. We showed that weak rigidity is a condition milder than distance rigidity for a framework, and derived a necessary and sufficient graphical condition for infinitesimal weak rigidity in the plane. The proposed graphical condition can easily verify infinitesimal weak rigidity of a framework without examining rank of the rigidity matrix, whereas no graphical conditions for infinitesimal rigidity exist in the literature. Two novel distributed formation control schemes via weak rigidity theory were also proposed. Our control strategies only require local relative displacement measurements, thus are distributed and communication-free. In particular, for the non-gradient based control law, local exponential stability of formation was obtained under a weakly rigid sensing graph. That is, our control law requires less information flowed in the network compared to the distance-based formation strategy, thus reduces costs and can be efficient in a more demanding environment. The future work includes:  1) the design of the gain matrix to stabilize the non-gradient based formation system, and  2) preservation of weak rigidity of the formation during agents' motion.

\section*{Acknowledgments} 
The authors would like to thank one of the reviewers for pointing out Reference \cite{Park17} in the second round of review.

\bibliographystyle{siamplain}
\bibliography{references}
\end{document}

%% file: ex_shared.tex
\usepackage{lipsum}
\usepackage{amsfonts}
\usepackage{graphicx}
\usepackage{epstopdf}
\usepackage{algorithmic}
\usepackage{color}
\ifpdf
  \DeclareGraphicsExtensions{.eps,.pdf,.png,.jpg}
\else
  \DeclareGraphicsExtensions{.eps}
\fi

\numberwithin{theorem}{section}

\newcommand{\TheTitle}{Weak Rigidity Theory and its Application to Formation Stabilization} 
\newcommand{\TheAuthors}{Gangshan Jing, Guofeng Zhang, Heung Wing Joseph Lee, and Long Wang}

\headers{Weak Rigidity Theory and Formation Stabilization}{G. Jing, G. Zhang, H. W. J. Lee, and L. Wang}

\title{{\TheTitle}\thanks{Submitted to the editors DATE.
\funding{This work was funded  in part by National Science Foundation of China (NSFC) grants (61751301 and 61533001) and Hong Kong RGC grants (531213 and 15206915).}}}

\author{
 Gangshan Jing\thanks{Center for Complex Systems, Xidian University, Xi'an 710071, China, and Department of Applied Mathematics, Hong Kong Polytechnic University, Hong Kong  (\email{nameisjing@gmail.com}).}
 \and
 Guofeng Zhang\thanks{Department of Applied Mathematics, Hong Kong Polytechnic University, Hong Kong (\email{Guofeng.Zhang@polyu.edu.hk},
    \email{Joseph.Lee@polyu.edu.hk}).}
  \and
 Heung Wing Joseph Lee\footnotemark[3]
 \and
 Long Wang\thanks{Corresponding author. Center for Systems and Control, College of Engineering, Peking University, Beijing 100871, China (\email{longwang@pku.edu.cn}).}
}
\usepackage{mathrsfs}
\usepackage{amssymb}
\usepackage{amsmath}
\allowdisplaybreaks[1]
\usepackage{amsfonts}
\usepackage{mathrsfs}\usepackage{latexsym, bm}
\usepackage{booktabs}
\usepackage{leftidx}
\usepackage{accents}
\usepackage{color}
\usepackage{qbordermatrix}
\usepackage{amsopn}

\newtheorem{remark}{Remark}
\newtheorem{problem}{Problem}
\newtheorem{assumption}{Assumption}
\newtheorem{example}{Example}